\documentclass[twocolumn]{aastex63}

\hypersetup{linkcolor=red,citecolor=blue,filecolor=cyan,urlcolor=black}

\newcommand{\OIII}{O\,{\scriptsize III}}
\newcommand{\OII}{O\,{\scriptsize II}}

\newcommand{\hst}{\textit{HST}}
\newcommand{\spitzer}{\textit{Spitzer}}
\newcommand{\jwst}{\textit{JWST}}
\newcommand{\eazy}{\texttt{EAzY}}
\newcommand{\lya}{Ly$\alpha$}

\usepackage{float}
\usepackage{amsmath}
\usepackage{graphicx} 
\usepackage[caption=false]{subfig}

\accepted{\today}
\submitjournal{ApJ}

\graphicspath{{./}{figures/}}

\shorttitle{The Physical Properties of Luminous $z\gtrsim$8 Galaxies}
\shortauthors{Roberts-Borsani et al.}

\begin{document}


\title{The Physical Properties of Luminous $z\gtrsim8$ Galaxies and Implications for the Cosmic Star Formation Rate Density From $\sim$0.35 deg$^{2}$ of (Pure-)Parallel \hst\ Observations
\footnote{Based on observations made with the NASA/ESA Hubble Space Telescope, obtained from the data archive at the Space Telescope Science Institute. STScI is operated by the Association of Universities for Research in Astronomy, Inc. under NASA contract NAS 5-26555.}}


\correspondingauthor{Guido Roberts-Borsani}
\email{guidorb@astro.ucla.edu}

\author[0000-0002-4140-1367]{Guido Roberts-Borsani}
\affiliation{Department of Physics and Astronomy, University of California, Los Angeles, 430 Portola Plaza, Los Angeles, CA 90095, USA}

\author[0000-0002-8512-1404]{Takahiro Morishita}
\affiliation{Space Telescope Science Institute, 3700 San Martin Drive, Baltimore, MD 21218, USA}

\author[0000-0002-8460-0390]{Tommaso Treu}
\affiliation{Department of Physics and Astronomy, University of California, Los Angeles, 430 Portola Plaza, Los Angeles, CA 90095, USA}

\author[0000-0003-4570-3159]{Nicha Leethochawalit}
\affiliation{School of Physics, University of Melbourne, Parkville 3010, VIC, Australia}
\affiliation{ARC Centre of Excellence for All Sky Astrophysics in 3 Dimensions (ASTRO 3D), Australia}

\author[0000-0001-9391-305X]{Michele Trenti}
\affiliation{School of Physics, University of Melbourne, Parkville 3010, VIC, Australia}
\affiliation{ARC Centre of Excellence for All Sky Astrophysics in 3 Dimensions (ASTRO 3D), Australia}

\begin{abstract}
We present the largest systematic, \hst-based search to date for luminous $z\gtrsim8$ galaxy candidates using $\sim$1267 arcmin$^{2}$ of (pure-)parallel observations from a compilation of 288 random sightlines with ACS and WFC3 observations, derived from the SuperBoRG data set and together representing a factor $\sim1.12\times$ larger than existing space-based data sets. Using NIR color cuts and careful photo-$z$ analyses, we find 31 $z\gtrsim8$ galaxy candidates over 29 unique sightlines, and derive global galaxy properties such as \textit{UV} magnitudes and continuum slopes, sizes, and rest-frame optical properties (e.g., SFRs, stellar masses, $A_{\rm v}$). Taking advantage of the (pure-)parallel nature of our data set - making it one of the most representative thus far - and derived SFRs, we evaluate the cosmic star formation rate density for the bright end of the \textit{UV} luminosity function at $z\sim8-10$ and test the validity of luminosity function-derived results using a conversion factor. We find our method yields comparable results to those derived with luminosity functions. Furthermore, we present follow up observations of 5 (Super)BoRG targets with Keck/MOSFIRE, finding no evidence of Ly$\alpha$ in $>3$ hrs of $Y-$band observations in either, consistent with a largely neutral medium at $z\sim8$. Our results offer a definitive \hst\ legacy on the bright end of the luminosity function and provide a valuable benchmark as well as targets for follow up with \jwst.
\end{abstract}


\section{Introduction}
Determining the primary drivers of cosmic reionization - where the Universe transitioned from a neutral state to a fully ionized one - remains one of the holy grails of modern extragalactic astrophysics. Such a process is thought to occur over the first billion years of the Universe (corresponding approximately to redshifts of $z>6$), thus making the detection and characterization of galaxies at those epochs a prerequisite. To this end, the last decade has seen remarkable progress in enlarging $z\gtrsim6$ galaxy samples to their thousands, in large part thanks to the highly sensitive near-infrared (NIR) capabilities of the \textit{Hubble} Space Telescope's (\hst) Wide Field Camera 3 (WFC3) and ground-based spectroscopic efforts, revealing hundreds of sources at redshifts of $7<z<8$ (e.g., \citealt{bradley12,bowler14,schmidt14,bouwens15,finkelstein15,laporte17,stefanon17}) and a dozen sources out to redshifts as high as $9<z<11$ (e.g., \citealt{coe13,ellis13,oesch14,oesch16,oesch18,hashimoto18,bouwens19,bowler20,laporte21}).

In particular, efforts to constrain the evolution of the faintest $z>7$ galaxies within the context of the \textit{UV} luminosity function (LF) have greatly benefited from observations of lensing clusters (e.g., the CLASH and Hubble Frontier Fields programs; \citealt{postman12} and \citealt{lotz17}, respectively) and deep observations over blank fields (e.g., the Hubble Ultra and eXtreme Deep Fields; \citealt{beckwith06} and \citealt{illingworth13}, respectively), however more moderate progress has been made in characterizing the evolution of the most luminous objects given their rarity and the especially large volumes required to identify them. Furthermore, such an enterprise is further limited by the availability of sufficiently deep, multi-band NIR data with which to identify the so-called "Lyman-break". Several programs have aimed to address these deficiencies through larger-area, multi-wavelength observations in legacy fields such as the UKIDSS UDS \citep{lawrence07}, UltraVISTA \citep{mccracken12} and the CANDELS programs \citep{grogin11,koekemoer11}, or through pure-parallel data sets such as the BoRG \citep{trenti11,bradley12,trenti12,schmidt14,calvi16,morishita18,rojasruiz20}, HIPPIES \citep{yan11} and ZFOURGE surveys \citep{tilvi13}, revealing several hundred $z>7$ luminous galaxy candidates that serve as ideal targets for spectroscopic follow up studies aiming to confirm their redshifts (e.g., \citealt{pentericci18}), characterize their rest-frame \textit{UV} properties (e.g., \citealt{finkelstein15,stark17,laporte17}), and/or constraining the opacity of the intergalactic medium (e.g., \citealt{treu13,mason18,hoag19}). Placing the interpretations of such results within the context of early galaxy evolution, however, remains even more challenging due to cosmic variance effects that are likely to plague single sightline observations and questions remain over how representative samples found over such fields are likely to be. To this end, the most unbiased samples come from independent sightlines of pure-parallel data sets, which provide uncorrelated observations over random patches of the sky and combined often afford similarly large search areas to the largest single-pointing legacy fields such as e.g., the CANDELS fields (while this is especially true for space-based surveys, ground-based surveys generally afford much larger areas of a few square degrees, albeit at often much shallower depths - see e.g., \citealt{stefanon19,bowler20}). In this paper, therefore, we present results using a subset of the Super- Brightest of Reionizing Galaxies (SuperBoRG) data set \citep{morishita21}, a compilation of extragalactic parallel programs with \hst\ which spans 316 independent sightlines and a total effective search area of $\sim0.41$ deg$^{2}$ with optical and near-IR imaging. The compilation of these data sets includes consistently reduced imaging with an updated data reduction pipeline of ACS (F435W, F475W), WFC3-UVIS (F300X, F350LP, F475X, F555W, F600LP, F606W, F625W, F775W, F814W, F850LP) and WFC3-IR (F098M, F105W, F110W, F125W, F140W, F160W) bands, enabling a large wavelength range over which to sample the rest-frame \textit{UV} spectrum of $z>8$ galaxies and exclude low redshift contaminants through the Lyman-break technique. In addition to \hst\ imaging, the SuperBoRG data set also includes overlapping \spitzer/IRAC 3.6 $\mu$m and 4.5 $\mu$m imaging over a significant portion ($\sim50$\%) of the data set, thus extending the full wavelength coverage from $\sim$0.3-1.6 $\mu$m to $\sim$0.3-4.5 $\mu$m and offering valuable constraints for $z\gtrsim8$ galaxy properties and the exclusion of low-$z$ contaminants. With a median (5$\sigma$) F160W depth and (1$\sigma$) field-to-field scatter of $\sim26.5\pm0.4$ AB (affording a $\sim0.1$ mag improvement compared to the originally-reduced data sets) and the large wavelength coverage afforded by the large assortment of optical-to-NIR filters, the survey and data sets utilized here are ideally suited to the search for luminous high redshift ($z\gtrsim8$) galaxies while the (pure-)parallel nature and large area probed ensure that derived samples are robust against the effects of cosmic variance and representative of the general galaxy populations at high redshift, allowing for definitive legacy constraints on the bright end of the luminosity function and providing ideal follow up targets for upcoming observations with \jwst\ (e.g., GO 1747, PI: Roberts-Borsani). In this paper, therefore, we conduct and present one of the largest and most unbiased space-based searches to date for $z\gtrsim8$ galaxy candidates with the aim of placing valuable constraints on their observed properties and providing prime targets for \jwst\ follow up. The paper itself is structured as follows. In Section \ref{sec:sampselect} we describe the adopted data sets as well our selection criteria for searching for high redshift candidates. We then discuss the physical properties of our derived $z\gtrsim8$ sample, including absolute magnitudes, \textit{UV} continuum slopes, galaxy sizes, and rest-frame optical properties in Section \ref{sec:sampleprops}, and describe follow up observations with Keck/MOSFIRE of several targets in Section \ref{sec:keck}. Making use of the derived rest-frame optical properties and the unbiased nature of our sample, we conclude our paper by conducting a unique exploration and comparison of assumptions used for derivations of the cosmic star formation rate density at $z\sim8-10$ and explore whether such assumptions hold at such high redshifts. We discuss this in Section \ref{sec:csfrd}, provide forecasts for \jwst\ in Section \ref{sec:jwst}, and finally summarize our findings in Section \ref{sec:conclusions}. Making use of the galaxy samples found in this study, derivations of the SuperBoRG \textit{UV} luminosity functions will be presented in a forthcoming paper (Leethochawalit et al. 2022, in prep.). Throughout this paper, we refer interchangeably to the \hst\ F098M, F105W, F125W, F140W,
and F160W bands as $Y_{098}$, $Y_{105}$, $J_{125}$, $JH_{140}$ and $H_{160}$, respectively, for simplicity. Where relevant, we assume \textit{H}$_{0}=$70 km/s/Mpc, $\Omega_{m}=$0.3, and $\Omega_{\wedge}=$0.7. All magnitudes are in the AB system \citep{oke83}.

\section{Sample Selection}
\label{sec:sampselect}
We begin by outlining the construction of the primary photometric data sets utilized in this study to search for luminous $z\gtrsim8$ galaxy candidates. For such a search, sufficient wavelength coverage in the optical and NIR are required to constrain the location of the Lyman break by means of NIR colors.


\subsection{\hst\ and \spitzer/IRAC Photometry}
\label{subsec:extractedphoto}
To construct our photometric data set, we make use of a subset of the SuperBoRG catalogs, which act as a compilation of several, homogeneously-reduced extragalactic \hst/WFC3 (pure-)parallel surveys. For full details on the SuperBoRG compilation and data reduction, we refer the reader to \citet{morishita20} and \citet{morishita21}, however we also provide a short description here, for convenience. Briefly, the compilation of \hst\ photometry includes pure-parallel observations from the BoRG survey (specifically, Cycles 17, 19, 22 and 25; \citealt{trenti11,bradley12,schmidt14,calvi16,rojasruiz20,morishita21}), Cycle 17 of the Hubble Infrared Pure Parallel Imaging Extragalactic Survey (HIPPIES; \citealt{yan11}) and the COS-GTO data set, as well as coordinated-parallel observations from the Cluster Lensing And Supernonva survey with Hubble (CLASH; \citealt{postman12}) and the Reionization Lensing Cluster Survey (RELICS; \citealt{salmon18,coe19}). Although included in the SuperBoRG data set, we do not include HIPPIES Cycle 18 observations due to most bands having single exposures per filter with limited temporal sampling of the detector (SPAR200), which leads to enhanced probability that cosmic rays and detector artifacts are present in the final reduced (science) images. Combined, the adopted data sets provide optical-to-NIR imaging through ACS (F435W, F475W), WFC3-UVIS (F300X, F350LP, F475X, F555W, F600LP, F606W, F625W, F775W, F814W, F850LP) and WFC3-IR (F098M, F105W, F110W, F125W, F140W, F160W).

While most (Super)BoRG fields were selected to lie at high Galactic latitudes ($|b|>30^{\circ}$) to avoid significant extinction effects or confusion by foreground stars, all photometry in the SuperBoRG data set was corrected for Galactic extinction along the line of sight, assuming a \citet{cardelli89} extinction law. The above \hst\ photometry was extracted using the \texttt{SExtractor} \citep{bertin96} tool in combination with a stacked WFC3/F140W+F160W detection image\footnote{\url{https://archive.stsci.edu/hlsp/superborg}}. Such a red image is ideally suited toward the detection of fainter $z\gtrsim9$ galaxies, since the Lyman break begins to drop out of the F125W band at those redshifts. However, for $z\sim8$ galaxies, where the Lyman break remains blueward of the F125W band, a F125W detection image can be better suited for source detections, since the generally blue \textit{UV} slopes of high-$z$ may render detections in the reddest filters challenging. Furthermore, we point out that the F160W filter generally affords lower signal-to-noise (S/N) than the F125W filter for the same exposure time of a flat spectral energy distribution (SED) in $F_{\nu}$. For these reasons, and to remain as complete as possible, we re-run \texttt{SExtractor} over the F125W images in the exact same manner as detailed in \citet{morishita20} in order to create F125W-derived photometric catalogs which will be publicly-released in a forthcoming data release. All detection images are convolved to the F160W resolution prior to being used with \texttt{SExtractor} and where duplication of sources (for $z\sim8$ dropout searches only) occurs we assume the F125W-detected photometry. S/N ratios are calculated from the flux and error measured in an aperture with a diameter size of $0.\!\arcsec64$. We then scale the measured fluxes of all filters by uniformly multiplying the ratio of total flux and aperture flux measured in the detection band. Limiting magnitudes for both detection bands are calculated by using the consistent detection setup for real sources, in the same way as presented in Section 3.1.3 of \citet[][]{morishita21}.

In some cases, the \hst\ observations of SuperBoRG fields overlap with observations taken with \spitzer/IRAC (140/288 fields, or $\sim$49\% of the search area), providing useful additional constraints with which to constrain the high-$z$ nature of candidates and their rest-frame optical properties. To extract the IRAC photometry, we adopt the same procedure described in \citet{morishita20}: given the lower spatial resolution of the IRAC images, we apply the \texttt{GALFIT} \citep{peng02,peng10} tool to the high-resolution \hst\ detection images in order to obtain the structural parameters (radius, S\'ersic index, axis ratio) of the high-$z$ sources and use these as fixed parameters to infer the galaxy profiles on the low-resolution IRAC images (magnitude and central position are left as free parameters). Neighbouring sources within a 150 pixel (i.e., 12 arcsec) radius from the high-$z$ source and with $>20$\% of its brightness are fitted simultaneously, while all other sources are masked.

\begin{table}
    \centering
    \begin{tabular}{lccc}
        \hline
        \hline
        Filter & $5\sigma$ Depth & $1\sigma$ Scatter & Nr. of fields \\
         & [AB] & [AB] & \\
        \hline
        \hst/F300X & 26.31 & 0.15 & 3 \\
        \hst/F435W & 25.13 & --- & 1 \\
        \hst/F475W & 25.87 & --- & 1 \\
        \hst/F475X & 26.36 & 0.42 & 5 \\
        \hst/F350LP & 26.74 & 0.27 & 184 \\
        \hst/F606W & 26.64 & 0.52 & 83 \\
        \hst/F600LP & 26.06 & 0.43 & 27 \\
        \hst/F814W & 25.39 & 0.41 & 54 \\
        \hst/F850LP & 24.65 & --- & 1 \\
        \hst/F098M & 26.71 & 0.43 & 120 \\
        \hst/F105W & 26.67 & 0.34 & 178 \\
        \hst/F110W & 26.77 & 0.47 & 13 \\
        \hst/F125W & 26.67 & 0.37 & 283 \\
        \hst/F140W & 26.64 & 0.40 & 184 \\
        \hst/F160W & 26.45 & 0.38 & 288 \\
        \spitzer/IRAC/CH1 & 23.88 & 0.30 & 131 \\
        \spitzer/IRAC/CH2 & 23.31 & 0.27 & 98 \\
        \hline 
    \end{tabular}
    \caption{The median $5\sigma$ depths and $1\sigma$ field-to-field scatter over all the utilized SuperBoRG fields (see Section \ref{subsec:extractedphoto} for details) in this study. The $1\sigma$ scatter refers to the field-to-field scatter rather than the intrinsic uncertainties over each filter.}
    \label{tab:filt_depths}
\end{table}

\begin{figure}
\center
 \includegraphics[width=\columnwidth]{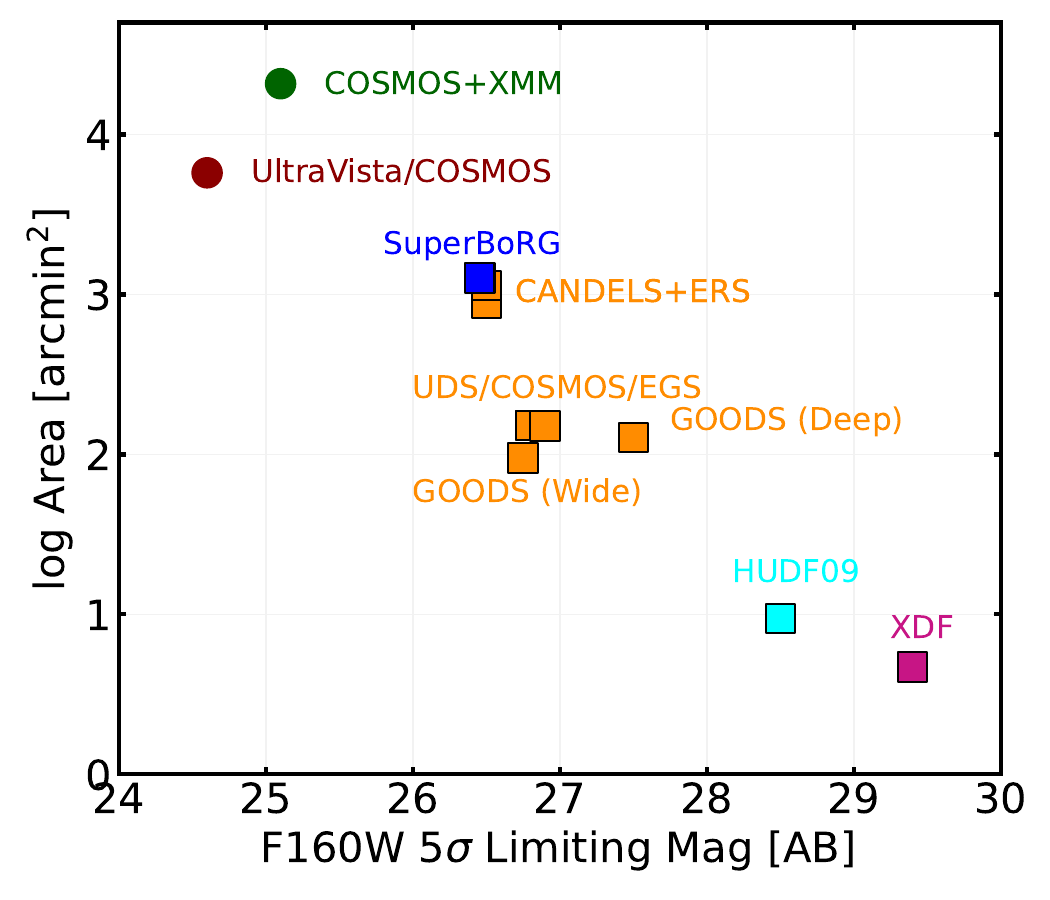}
 \caption{The median depth and total area of the SuperBoRG data utilized in this work (blue square), compared to those of other space-based surveys (squares): CANDELS (orange, where GOODS-North and GOODS-South are combined), HUDF09 (1+2; cyan) and XDF (magenta). The UDS, COSMOS and EGS labels are plotted as one given the proximity of each of their associated squares. All literature points are taken from \citet[Table 1]{bouwens15} and \citet{bouwens19} (all CANDELS fields+ERS) and serve as an approximate comparison: our combined data set affords approximately 1.4$\times$ more area than the CANDELS+ERS search area. For reference, ground-based searches from \citet{stefanon19} and \citet{bowler20} over the COSMOS+XMM (green circle) and UltraVista/COSMOS (dark red circle) fields, respectively, are also plotted (median depths and total areas) to illustrate the larger areas but shallower depths of ground-based surveys.}
 \label{fig:depth}
\end{figure}

Overall, the combined data set described above probes 288 independent sightlines, spanning an effective search area of $\sim$1267 arcmin$^{2}$, which is a factor $\sim1.12-1.4\times$ greater than the largest space-based searches using extragalactic legacy surveys (i.e., $\sim883$ arcmin$^{2}$ presented in \citealt{bouwens19} using CANDELS-WIDE plus an additional 147 arcmin$^{2}$, or $\sim$1136 arcmin$^{2}$ from the combination of the five CANDELS fields, Frontier Field parallel pointings, and the Hubble Ultra Deep Field as presented in \citealt{bouwens21}). The total SuperBoRG search area utilized in this work and median depth(s) are tabulated in Table \ref{tab:filt_depths} and plotted in Figure~\ref{fig:depth}, along with the various \hst\ survey depths and areas presented in \citet{bouwens15}, \citet{bouwens19}, \citet{stefanon19} and \citet{bowler20}. Perhaps even more importantly, the (pure-)parallel nature of the compilation means the effects of cosmic variance and galaxy clustering are reduced and thus offers a unique means with which to conduct an unbiased selection and study of $z\gtrsim8$ galaxy evolution.

\subsection{Color Selection and Visual Inspection}
\label{sec:colors}
To derive a robust sample of $z\gtrsim8$ galaxy candidates, we begin by utilizing color selection criteria similar to those adopted by several previous BoRG analyses \citep{bradley12,schmidt14,calvi16}, and described in detail by \citet[see Section~5.2]{morishita21}. The selections are based on the Lyman-break technique and include additional S/N and NIR color cuts to ensure the exclusion of low redshift interlopers. The Lyman-break color cuts that we employ are the following. For $z\sim8$ galaxy candidates where observations with the $Y_{\rm 098}$ filter are available, we require

$$S/N_{J_{\rm125}}>6.0 \,\,\land$$
$$S/N_{H_{\rm 160}}>4.0 \,\,\land$$
$$Y_{098}-J_{\rm125}>1.75 \,\,\land$$
$$J_{\rm125}-H_{\rm 160}<0.5 \,\,\land$$
$$(J_{\rm125}-H_{\rm 160})<0.02+0.15\cdot(Y_{098}-J_{\rm125}-1.75)$$

and non-detections (S/N$<$1.0) in the $I_{814}$ band and all filters at bluer wavelengths. For fields where observations with the $Y_{\rm 105}$ filter are available, we require

$$S/N_{J_{\rm125}}>6.0 \,\,\land$$
$$S/N_{H_{\rm 160}}>4.0 \,\,\land$$
$$Y_{105}-J_{\rm125}>0.45 \,\,\land$$
$$J_{\rm125}-H_{\rm 160}<0.5 \,\,\land$$
$$Y_{105}-J_{\rm125}>1.5\cdot(J_{\rm125}-H_{\rm 160})+0.45$$

and identical non-detections in the filters used for $Y_{\rm 098}$ dropouts. The satisfaction of either or both of the above selection criteria is sufficient for the determination of $z\sim8$ source, and we do not separate between the two in our final samples since such distinction is not relevant for our analyses. Given the overlap between the $Y_{\rm 098}$ and $Y_{\rm 105}$ filters, however, we note and reassure the reader that no case occurs where a galaxy satisfies the $Y_{\rm 098}$ criteria but not the $Y_{\rm 105}$ criteria. For $z\sim9$ dropouts, we utilize the $JH_{\rm 140}$ band instead, given the redder wavelengths that the Lyman-break is stretched to. The criteria are

$$S/N_{JH_{\rm 140}}>6.0 \,\,\land$$
$$S/N_{H_{\rm 160}}>4.0 \,\,\land$$
$$Y_{105}-JH_{\rm 140}>1.5 \,\,\land$$
$$JH_{\rm 140}-H_{\rm 160}<0.3 \,\,\land$$
$$Y_{105}-JH_{\rm 140}>5.33\cdot(JH_{\rm 140}-H_{\rm 160})+0.7$$

and we require non-detections in all filters bluer than $Y_{\rm 105}$. Finally, for our $z\sim10$ dropouts, we require

$$S/N_{H_{\rm 160}}>6.0$$
$$J_{\rm125}-H_{\rm 160}>1.3$$

as well as non-detections in $Y_{\rm 105}$ and all bluer filters. Furthermore, for all resulting $z\sim8-10$ dropout candidates where \spitzer/IRAC 3.6 $\mu$m coverage is available, we require an additional color cut of $H_{\rm 160}-[3.6]<1.4$. This latter consideration ensures we take advantage of the filter's strong ability to distinguish between low redshift interlopers (see \citealt{bouwens15}). Finally, we also remove candidates with a \texttt{SExtractor} stellarity parameter greater than 0.95 to avoid confusion with nearby stars. The $z\sim8$ selection is performed on both the F125W and F140W+F160W catalogs for completeness, while the $z>8$ selection is performed on the F140W+F160W catalog only due to it being better suited to detecting galaxies at those redshifts.

\begin{figure*}
\center
 \includegraphics[width=\textwidth]{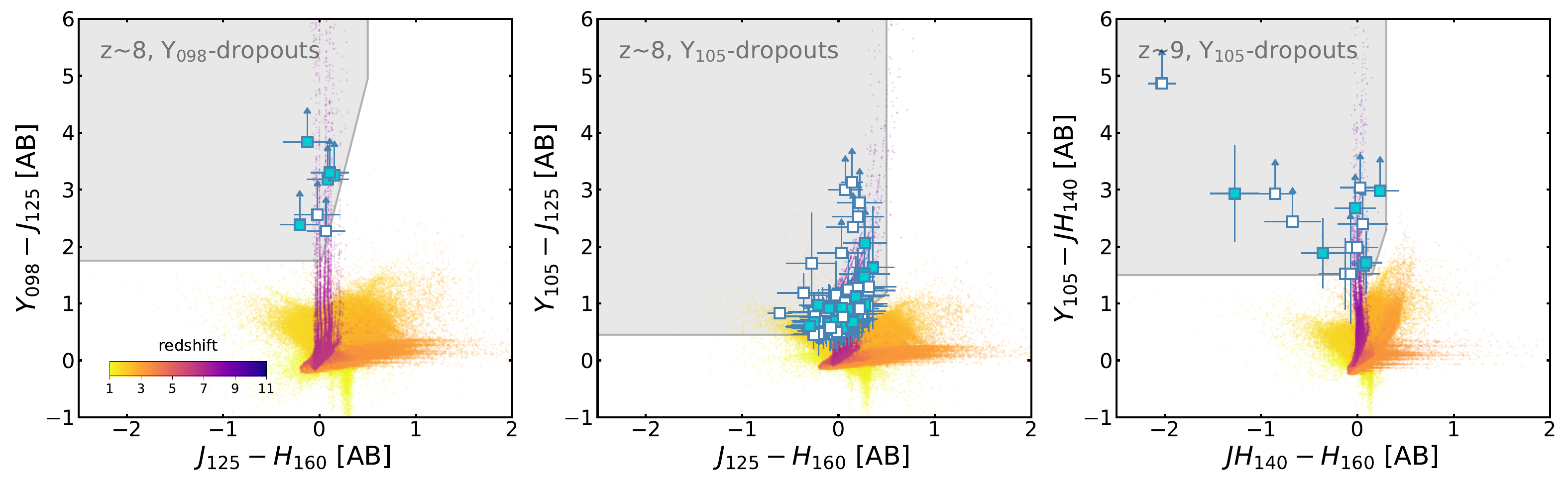}
 \caption{The NIR colors of our fiducial sample of $z\sim8$, $z\sim9$ and $z\sim10$ SuperBoRG dropout galaxies (squares). Each galaxy is color-coded according to whether the resulting photo-$z$ from \eazy\ agreed with a $z>7.5$ (blue fill) or $z<7.5$ (white fill) solution. The colors of mock $0<z<15$ galaxies over a single realization of the EGG simulations (see Section \ref{subsubsec:prior} for details) are overplotted, for comparison.}
 \label{fig:highzcolors}
\end{figure*}

Upon selection, all high-$z$ candidates are visually inspected by two of the authors (GRB and NL/TM). All candidates are flagged an integer of 0 (clearly not a dropout), 1 (clear dropout) or 2 (suspicious). For a 1 flag, a candidate must not (i) be affected by a partial lack of data (i.e., sources close to the detector edge) or (ii) have its IR bands impacted by persistence or clear artifacts (e.g., streaks). As an additional safety check to asses photometric quality and maintain a high quality dropout sample, we further inspect the images to ensure the object displays a clear dropout nature from the IR to optical bands, and no significant flux in any of the individual or stacked optical bands. If a candidate displays any of the aforementioned artifacts or significant optical flux in any of the individual or stacked optical images, it is flagged as a clear non-dropout (i.e., 0). Galaxies where a very small degree of optical flux in several (or all) optical bands may appear to be present but not necessarily indicative of an optical detection (e.g., due to high fluctuations in the surrounding noise or hot pixels), or whose isolated nature is somewhat unclear (e.g., if the object resides near a particularly bright and/or large object or close to an extended object) are allocated a flag of 2, since their dropout nature is unclear but not to the degree that it warrants exclusion. The dropout samples are then divided into two sub-samples: a high confidence or flagged sample. A high confidence galaxy requires two 1 flags from the two authors, while a galaxy with both a 1 and 2 flag is placed in the flagged sample. Galaxies with two 2 flags or a single 0 flag are discarded. Assuming the highest redshift dropout bin in the case where several criteria are satisfied, our selection yields a total of 206 $z\sim8$ dropouts (108 high confidence and 98 flagged), 17 $z\sim9$ dropouts (14 high confidence and 3 flagged) and 37 $z\sim10$ dropouts (15 high confidence and 22 flagged), over 110, 15 and 32 unique and independent sightlines, respectively. As will be discussed in Section \ref{subsubsec:prior}, the larger number of $z\sim10$ dropouts compared to $z\sim9$ dropouts can likely be attributed to a higher contamination fraction from low-$z$ interlopers. The selection colors of the high confidence dropouts are shown in Figure~\ref{fig:highzcolors}, while we offer a comparison of our dropout selection with previously selected BoRG galaxies in Appendix \ref{sec:compare}.

\subsection{Photometric Redshift Estimation}
\label{subsec:photoz}
\subsubsection{Construction of an empirical prior}
\label{subsubsec:prior}
While the selection of high-$z$ candidates based on NIR colors is considered standard practice, the probability of low-$z$ interlopers can be further reduced by additional steps. To ensure our dropout galaxies are truly at high redshift, we refine our high-$z$ sample selection through photometric redshift estimates with \eazy\ \citep{brammer08}. Given the relatively limited photometric information available (due to sensitivity and wavelength coverage of \hst/\spitzer), and the very low density in the sky of galaxies at $z\sim8$ with respect to foreground contaminants \citep{vulcani17}, an informative prior is essential to obtain a meaningful posterior distribution function for the photo-$z$ \citep{stiavelli09}.

\begin{figure}
\center
 \includegraphics[width=0.75\columnwidth]{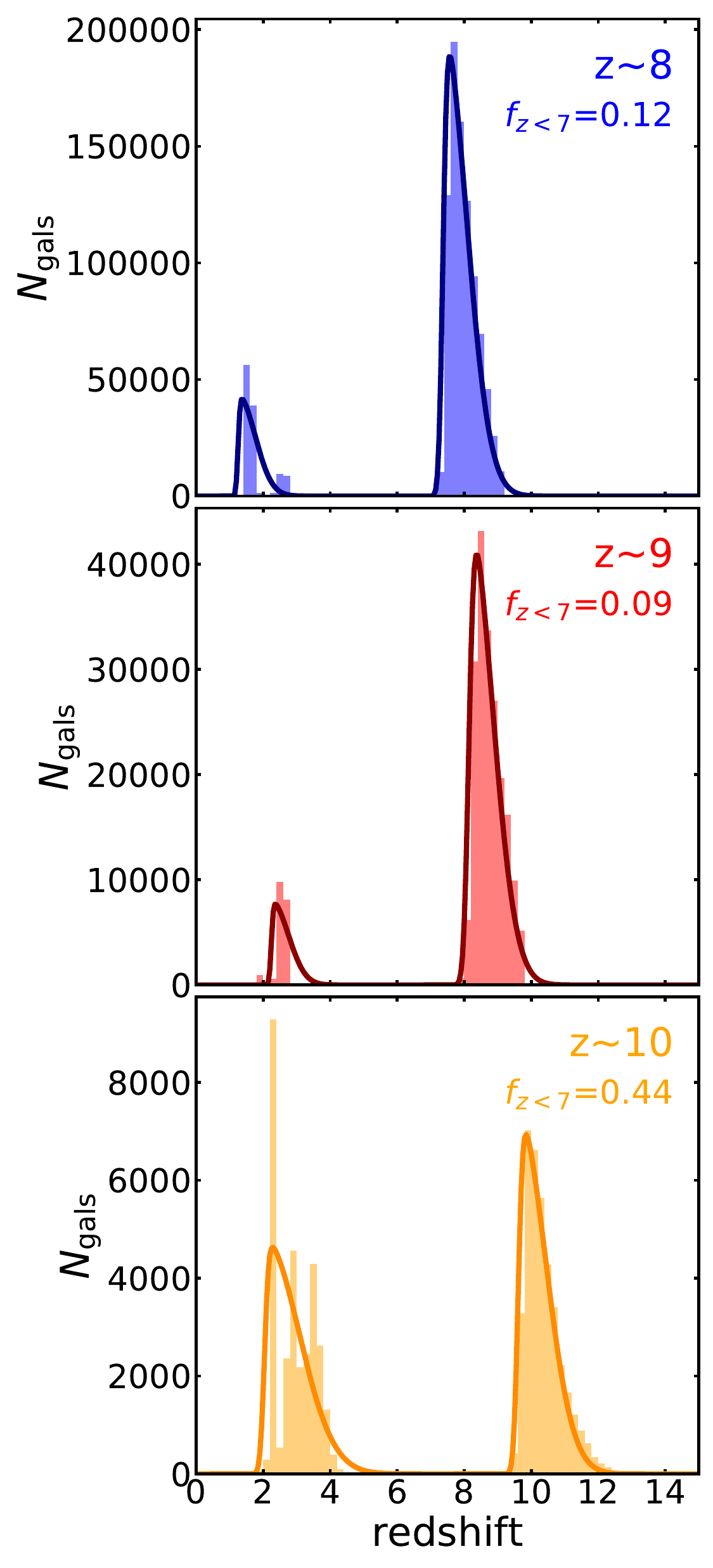}
 \caption{The distribution of redshifts belonging to the $z\sim8$, $z\sim9$ and $z\sim10$ EGG-simulated dropouts, using the NIR color cuts and S/N requirements described in Sections \ref{sec:colors} and \ref{subsubsec:prior}. In general, one observes a primary peak of high-$z$ galaxies resulting from the selection, as well as a more minor contribution from $z\sim1-4$ interlopers. The dark lines represent example fits of our double Skew-normal profiles, which account for multiple peaks of low-$z$ interlopers as well as the skewed distribution of high-$z$ galaxies.}
 \label{fig:prior}
\end{figure}

To construct such an informative prior, we make use of the Empirical Galaxy Generator (EGG) tool \citep{schreiber17} to generate representative mock galaxy photometry. The code makes use entirely of empirical prescriptions to match $0<z<7$ galaxy observations and generates its mock galaxy SEDs and photometry through (i) drawing redshifts and stellar masses from observed galaxy stellar mass functions, (ii) deriving star formation rates (SFRs) from the so-called galaxy ``main sequence'' and (iii) attributing dust attenuation, optical colors and disk/bulge morophologies from empirical relations derived from \hst\ and \textit{Herschel} observations of the CANDELS fields. Clustering effects are also simulated. We thus make use of the default set of \hst\ ACS and WFC3 bands to generate mock galaxy photometry over 10000 realizations of a 0.1 deg$^{2}$ field, providing $\sim1.4\times10^{8}$ simulated star-forming and quiescent galaxies over redshifts $0<z<15$ and stellar masses log\,$M_{*}/M_{\odot}>6$ (and down to F160W magnitudes of 28 AB). We subsequently scale the flux uncertainties from the median 5$\sigma$ limiting magnitudes of the SuperBoRG bands over the entire data set, prior to applying the NIR and S/N cuts described in Section \ref{sec:colors} to select the simulated dropouts. We note that while nearly all of the SuperBoRG fields are covered by virtually all of the WFC3 IR bands, the same cannot be said for the uniformity of the blue ACS and UVIS filters in each field. Thus, we opt to simplify the simulated observations by adopting from the default set of EGG filters those that are most prominent over the SuperBoRG fields, namely F606W and F814W. All other \hst\ NIR filters used in the selection criteria (i.e., F098M, F105W, F125W, F140W and F160W) are included, however we do not include \spitzer/IRAC photometry given the majority of our drouput sources either do not have such photometry or do in the form of mostly upper limits. Additionally, although in this study we make use of two detection images for $z\sim8$ dropouts, we make the assumption that the number densities do not change significantly due to the choice of NIR detection band over the number of realizations we perform and thus select galaxies based on their $H-$band magnitudes. The redshift distributions of the resulting $z\sim8-10$ dropouts (over all magnitudes and combining the $Y_{\rm 098}$ and $Y_{\rm 105}$ results for $z\sim8$) are plotted in Figure~\ref{fig:prior}.

Examining the results first over all magnitudes, for the simulated $Y_{\rm 098+105}-J_{\rm 125}$ $z\sim8$ dropouts, we observe a dominant peak of star-forming galaxies centered at $z\sim8$, with two secondary and smaller peaks at $z\sim1.5$ and $z\sim2.5$. The former consists primarily of quiescent-type galaxies while the latter of star-forming galaxies. If we define a low-$z$ interloper fraction, $f_{\rm int}$, as the number of dropout galaxies with a redshift below 7, we estimate a contamination fraction of 12\%. This fraction decreases to 9\% for $Y_{\rm 105}-JH_{\rm 140}$ $z\sim9$ dropouts, where we observe a well-defined peak centered at $z\sim9$ and a much smaller peak again at $z\sim2.5$ consisting entirely of star-forming galaxies. Finally, in our EGG simulations we observe a number of bright, $z\sim10$ candidates from $J_{\rm 125}-H_{\rm 160}$ NIR cuts, with a well-defined peak of star-forming galaxies found at the desired redshift, but equally find an important and somewhat broad contribution from both star-forming and quiescent low-$z$ galaxies at redshifts of $z\simeq2-4$. The fraction of contaminants is at its highest out of the three samples, with an estimated 44\% contamination rate. For comparison purposes only, we note that if were to include \spitzer/IRAC constraints in the simulations (namely, $H_{\rm 160}-[3.6]<1.4$), such contamination levels drop to 11\%, 6\% and 26\% for the $z\sim8$, $z\sim9$ and $z\sim10$ dropout samples. In similar fashion, if we were to adopt the F350LP filter as the only optical filter (which is the case over a non-negligible number of our fields), we find the contamination rate would remain largely unchanged. Additionally, we also note that since most of the low redshift interlopers in all dropout samples reside at $z\sim2-4$, a more conservative threshold of $z<4$ results in virtually identical contamination fractions. 

Similarly to \citet{morishita18}, to construct the prior itself we fit the redshift-magnitude distributions for each dropout sample with an analytical function. While \citet{morishita18} opt for a double-peaked Gaussian function, we instead opt for a double Skew-normal fit. The reason for this is that at low redshift two small peaks with differing amplitudes are often observed - one of star-forming galaxies at $z\sim1.75$ and one of quiescent galaxies at $z\sim2.25$ - and thus a standard Gaussian distribution often completely misses one of the two, while at high redshift the distribution often displays an abrupt decline at the lower-$z$ edge but a tail of objects at the higher-$z$ edge. This is particularly evident for the $z\sim10$ distribution). For this latter consideration, a Gaussian function would therefore overestimate the number of objects at the lower redshift edge and underestimate the number of objects at the higher redshift edge. A double Skew-normal function is thus a more appropriate fit, where tails of low-$z$ and high-$z$ objects can be accounted for by a non-zero skewness parameter while the function has the flexibility to return to standard double Gaussian fit if the skewness parameter is zero. We thus proceed to fit the redshift-magnitude distributions and normalize the resulting fit such that the peaks of the two Skew-normals sum to one. The (un-normalized) results of the procedure are presented in Table~\ref{tab:prior}.

\subsubsection{Photo-$z$ candidates}
Making use of the informative prior constructed in the previous section, we thus proceed with deriving photometric redshifts with \eazy. We adopt the default set of galaxy SED templates (v1.3) which have been shown to provide the least biased estimate for high-$z$ galaxies \citep{brinchmann17}. The templates also include SEDs of dusty, $z\sim2$ star-forming galaxies, as well as high equivalent width (EW) emission lines, which can significantly impact the observed photometry. We adopt an allowed redshift range of $z\in[0.01,15]$ in steps of $\Delta z=0.01$. Additionally, to ensure the redshift determinations remain as independent as possible from the assumed star formation histories (SFHs) of the templates, we allow a linear combination of all templates simultaneously. For all other options we adopt the default \eazy\ settings. 

The final photometric redshift is then determined via two criteria. The first is that the best-fit redshift ($z_{p}$) lies between $z_{p}=7.5-8.5$, $z_{p}=8.5-9.5$, and $z_{p}=9.5-10.5$ for $z\sim8$, $z\sim9$ and $z\sim10$ samples, respectively, and so on and so forth for even higher redshift candidates (although not explicitly required, we note that all $z_{\rm phot}>10.5$ candidates satisfy only the $z\sim10$ dropout criteria). The second is that the summed $z>6.5$ posterior distribution ($P(z\rm>6.5)$, normalized by the sum of the distribution across the entire redshift range) is greater than 0.7 for $z\sim8$ galaxies and that the summed $z>7.5$ posterior distribution is greater than 0.7 for $z\sim9-12$ galaxies. This latter consideration guards against selecting false-positive high-$z$ candidates based on poorly constrained redshift posterior distributions, where the choice of $z_{p}$ as the best-fit redshift is uninformative and the likelihood of selecting low-$z$ interlopers is increased. 

The photo-$z$ estimation combined with our $P(z)$ requirements reduces the high-confidence dropout samples to 35, 6, 2, galaxies at $z\sim8$, $z\sim9$, $z\sim10$, respectively (over 39 unique fields spanning a total area of $\sim$172 arcmin$^{2}$). Similarly, for our flagged sample of dropouts the sample sizes reduce to 23, 1, 4, 3 and 2 candidates for $z\sim8$, $z\sim9$, $z\sim10$, $z\sim11$ and $z\sim12$ redshift bins (over 31 unique fields spanning an effective area of $\sim$136 arcmin$^{2}$). We briefly note that, as shown in Figure~\ref{fig:prior}, some (expected) overlap between dropout samples is allowed by the photometric redshift prior, meaning that galaxies selected at a particular dropout redshift may end up in a different redshift bin according to their final photometric redshift. As a sanity check, we verify that no cases are found where transfer between redshift bins would contradict the observed photometry - e.g., that no $z\sim8$ dropout ends up in a $z>9.5$ photometric redshift bin or vice versa. No such cases were found, indicating that our adopted procedure for redshift determination is robust.

\begin{figure*}
\center
 \includegraphics[width=0.9\textwidth]{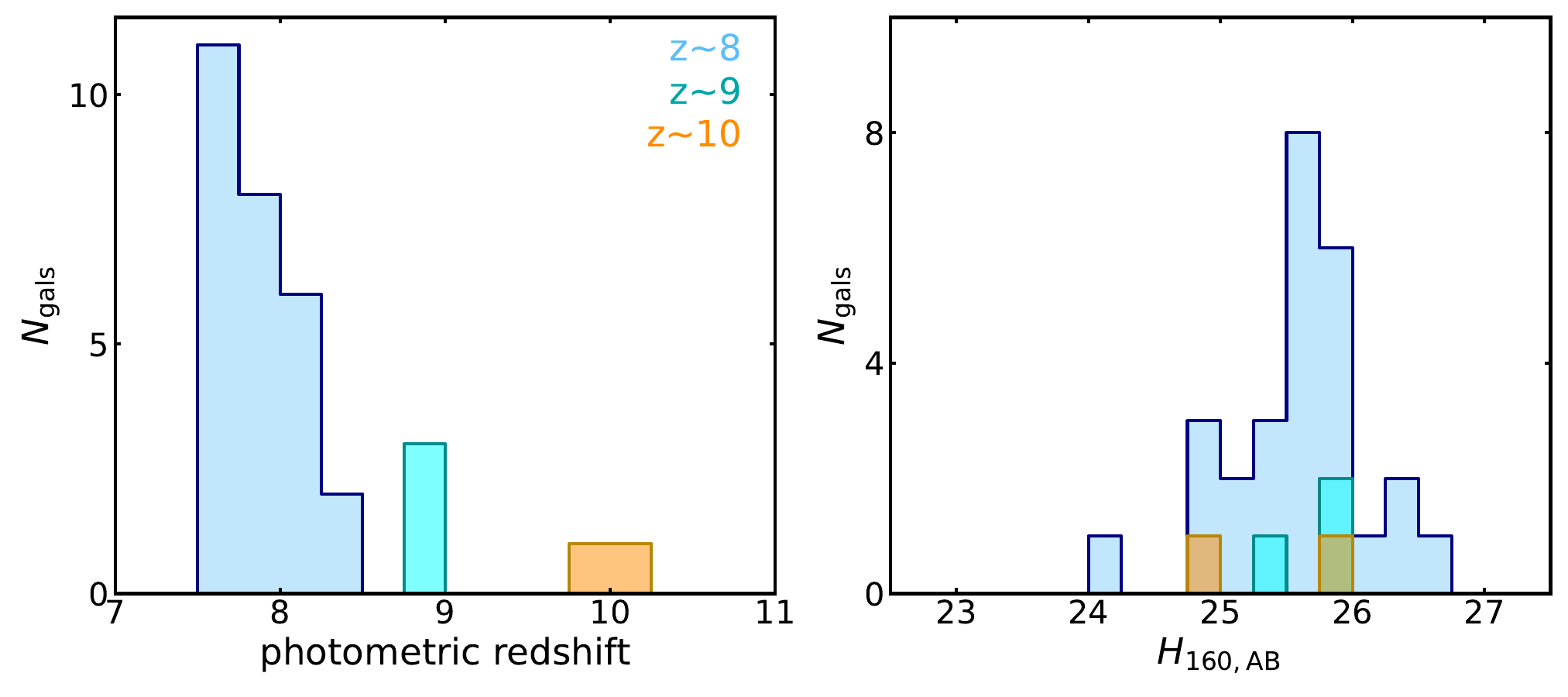}
 \caption{Left: The distribution of \eazy\ photometric redshifts for our fiducial $z\gtrsim8$ samples, in steps of $\delta z=0.25$. Right: The distribution of $H_{\rm 160}-$band magnitudes for the same galaxies, in steps of $\Delta m_{\rm 160}=0.25$ mag.}
 \label{fig:photozs}
\end{figure*}

\subsubsection{Accounting for spurious hot pixels}
The nature of pure-parallel imaging can, due to the lack of dithering, introduce spurious hot pixels that contaminate galaxy photometry and must be accounted for. For our fiducial sample of galaxies, therefore, we investigate the underlying RMS maps for the $Y-$ and $J-$bands (where the Lyman break is located at our redshifts of interest) for hot pixels. Upon inspection of the high-confidence photo-$z$ sample derived in the previous section, we find evidence for hot pixels in 5 galaxies (all $z\sim8$ dropouts; 0846$+$7654\_193, 1044$+$0349\_803, 1149$+$0133\_194, 1015$-$4714\_404, and 1102$-$3448\_365) and redefine the photometry by lowering the adopted count threshold to exclude such pixels. While this does not impact the color search of galaxies presented in this study, since our selection makes use of each \hst\ pointing's limiting magnitude in the case of a non-detection (e.g., in blue ACS filters or the $Y-$band) and thus is less susceptible to local artefacts, such a change could impact the photo-$z$ and $P(z)$ estimates. We thus rerun all of our \eazy\ fitting as described previously to establish the high-$z$ nature of each galaxy with the new photometry. With the new photometry, 4 of the candidates maintain their $z\gtrsim8$ nature, while the remaining galaxy (0846$+$7654\_193) has a new photo-$z$ solution of $z_{\rm phot}=1.47$, thereby reducing the sample to 42 high-confidence galaxies at $z\gtrsim8$.

\subsubsection{Possible sample contamination}
\label{subsubsec:contamination}
Some contamination of our photo-$z$ sample is to be expected. The primary culprits as discussed in Section \ref{subsubsec:prior} are bright, low redshift galaxies at $z\sim2-4$ that contribute significantly to NIR colors through large levels of dust or strong emission lines. Our simulations of NIR dropout galaxies suggests such contamination in $z\sim8-9$ samples is moderate (9-19\%), however the contamination in $z\sim10$ dropouts becomes much more significant at 44\%. This is especially relevant since we identify 11 $z\simeq10-12$ sources (2 high-confidence and 9 flagged based on visual inspection) over the full $\sim$1267 arcmin$^{2}$ search area (c.f. \citealt{oesch18} and \citealt{bouwens19}). Of these, 7 have \spitzer/IRAC constraints, which - as discussed in the previous section - greatly aids in distinguishing with low-$z$ interlopers. Furthermore, an additional consideration is contamination by brown dwarfs (in particular M-, L- and T-dwarfs), which can mimick the NIR photometry of bright, high redshift galaxies (see e.g., the detailed discussion by \citealt{bowler20}). While such contamination is expected to be relatively low at the redshifts we probe \citep{morishita20}, some minor contributions remain possible as we have thus far only limited the exclusion of point sources to the \texttt{SExtractor} stellarity parameter cut (i.e., exclude sources with stellarity $>0.95$). To mitigate any possible contamination as much as possible, we again make use of \eazy\ to fit each of our photo-$z$ candidates with the entire SpeX prism library\footnote{\url{http://pono.ucsd.edu/~adam/browndwarfs/spexprism/index.html}} of M-, L- and T-dwarf templates (one by one, 542 templates, extending from 0.65-2.55 microns; \citealt{burgasser14}) and compare the resulting $\chi^{2}$ with that of the best-fit galaxy spectrum. 7 $z\sim8$ and 3 $z\sim9$ galaxy candidates show slightly better $\chi^{2}$ values from the best-fit brown dwarf template, suggesting their high-$z$ nature cannot be reliably inferred. We exclude all galaxies where such a scenario applies.
An additional 5 $z\sim8$ galaxies from the flagged sample are also removed due to preference for a brown dwarf template, thus reducing our high-confidence and flagged samples to 32 and 28 $z\gtrsim8$ galaxies, respectively.

Additionally, for our fiducial and flagged samples of galaxies and added statistical robustness in subsequent analyses, we further derive a probability statistic for the galaxy nature of each object, based on the log-likelihood of the \eazy\ fits (galaxy and brown dwarf) and a prior. For the former, we use a Bayesian Information Criterion (BIC, defined as BIC=$k$ln($n$)$-$2ln($\hat{L}$), where $k$ is the number of free parameters, $n$ is the number of data points, and $\hat{L}$ is the log-likelihood), which assesses the goodness-of-fit of the best-fit model while effectively penalising the use of a large number of free parameters - e.g., in the case of the brown dwarf fit there is only one free parameter (i.e., the normalization of the SED) while for the galaxy fit (low- or high-$z$) there are 10 (9 amplitudes for each of the templates fit simultaneously and an additional redshift parameter). For the prior, we use the on-sky surface density of galaxies and brown dwarfs: for galaxies, we use the 32 high-confidence candidates found here over $\sim$1267 arcmin$^{2}$, while for brown dwarfs we use recent results from \citet{aganze21}, who found a total of 164 ultracool dwarfs in 0.6 deg$^{2}$ of spectroscopic data from \hst/WFC3 as part of the WFC3 Infrared Spectroscopic Parallel Survey \citep{atek10} and 3D-HST survey \citep{momcheva16}. More specifically, for the brown dwarf constraints we use only the 3D-HST results from \citet{aganze21}, where sufficient photometric coverage (i.e., WFC3/F606W, WFC3/F814W, WFC3/F125W, WFC3/F140W and WFC3/F160W) allows us to apply the same color-cuts to each object as adopted for each of our 32 high-$z$ candidates, in order to establish the number of brown dwarfs that could contaminate our samples. We thus apply each of the color cuts described in Section \ref{sec:colors} (with the exception of those requiring $Y-$band photometry, which is not available from 3D-HST). The number of brown dwarfs resulting from the SuperBoRG color cuts (2/51) over $\sim$600 arcmin$^{2}$ are then scaled to the 1257 arcmin$^{2}$ surface area to get the total contaminating number. Using the derived BICs and priors, our probability metrics can then be described as 

\begin{equation}
\begin{aligned}
P_{\rm gal} = \frac{\text{BIC}_{\rm BD}}{\text{BIC}_{\rm gal}}\times\frac{\text{Prior}_{\rm gal}}{\text{Prior}_{\rm BD}} \\
P_{\rm BD} = \frac{\text{BIC}_{\rm gal}}{\text{BIC}_{\rm BD}}\times\frac{\text{Prior}_{\rm BD}}{\text{Prior}_{\rm gal}}
\end{aligned}
\end{equation}

where $P_{\rm gal}$ is normalized to unity by the sum of both probability statistics (i.e., $P_{\rm gal}+P_{\rm BD}$) and tabulated in Table \ref{tab:candidates} and Table \ref{tab:candidates2}. Although not tabulated, $P_{\rm BD}$ is simply the remainder of $1-P_{\rm gal}$. We note that one candidate galaxy, 0037$-$3337\_669, has a $P_{\rm gal}$ value of 0.4 suggesting the source is more likely to be a brown dwarf than a galaxy and move it to our flagged sample. For illustration purposes, the best-fit brown dwarf templates for all remaining candidates are plotted as magenta lines in Figure \ref{fig:seds}. The large number of galaxy candidates found here (31 high-confidence galaxies) represents one of the largest samples of uniformly-derived $z\gtrsim8$ galaxies to date, and we consider this our fiducial sample for the rest of the paper. The distribution of the galaxies' photo-$z$s and $H-$band magnitudes are illustrated in Figure~\ref{fig:photozs}, while the main NIR colors and photo-$z$ properties from \eazy\ for the fiducial sample are presented in Appendix \ref{sec:primesample} and for the flagged sample in Appendix \ref{sec:flagsample}.

\section{The Physical Properties of $z\gtrsim8$ Galaxies}
\label{sec:sampleprops}
With our primary sample of high-$z$ galaxies now established, it is informative to infer and derive their global properties. As detailed in Section \ref{subsec:photoz}, we allow for a linear combination of all \eazy\ templates simultaneously, to avoid influence from the chosen SFH. However, such an approach does not guarantee a physically meaningful SFH, especially in the redder portion of the rest-frame UV and at the rest-frame optical where, in the absence of photometry longward of $H_{\rm 160}$ (e.g., \spitzer/IRAC data), the SED is often completely unconstrained. To provide more realistic SED fits and physically-motivated SFHs for our primary sample of galaxies, from which to derive global galaxy properties, we make use of the SED-fitting code \texttt{Bagpipes} \citep{carnall18} and fit the photometry of each galaxy candidate assuming a declining SFH at the high-$z$ redshift solutions from \eazy. The advantage of this approach is that the the best-fit SED solution is derived entirely from a nested sampling of the posterior distribution of best-fit galaxy parameters (within expected ranges), rather than a $\chi^{2}$ of a combination of galaxy SED templates which inevitably places greater weight on the \textit{a priori} choice of templates. The free parameters adopted and their ranges are listed in Table~\ref{tab:params}, while the best-fit SEDs are illustrated in Figure~\ref{fig:seds} and the postage stamp images for each galaxy shown in Figure~\ref{fig:postagestamps}.


\begin{table}
    \centering
    \begin{tabular}{lc}
        \hline
        \hline
        Parameter & High-$z$ fit\\
        \hline
        Redshift & Fixed to EAzY\\
        log stellar mass formed (M$_{\odot}$) & [$6,10$]\\
        Max. stellar age (Gyr) & [$0.001,1$]\\
        Metallicity (Z$_{\odot}$) & [$0,1$]\\
        log $U$ & [$-4,-1$]\\
        $A_{\rm v}$ (mag)$^{a}$ & [$0,1$]\\
        $\tau$ (Gyr) & [$0.001,100$]\\
        \hline 
    \end{tabular}
    \caption{The free parameters and allowed ranges in the declining star formation history model assumed with \texttt{Bagpipes} for a high-$z$ fit, used to generate realistic star formation histories with which to derive global galaxy properties. The high-$z$ fit assumes the redshift from \eazy. \\
    $^{a}$Assuming a \citet{calzetti2000} extinction law.}
    \label{tab:params}
\end{table}


\subsection{\textit{UV}-continuum slopes}
\label{subsec:UVslope}
\begin{figure*}
\center
 \includegraphics[width=0.8\textwidth]{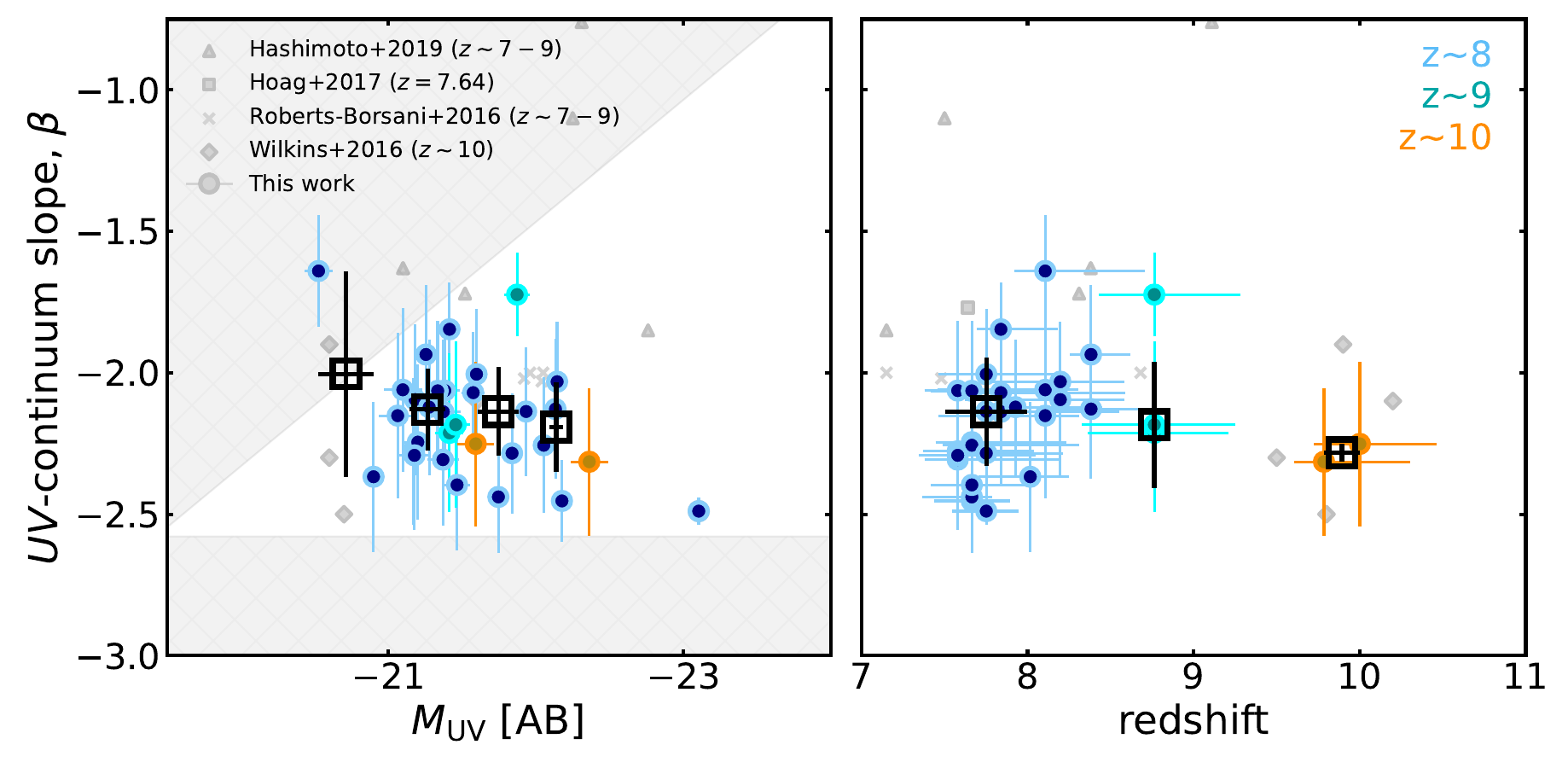}
 \caption{The \textit{UV} slopes and intrinsic magnitudes of our fiducial sample of $z\gtrsim8$ SuperBoRG galaxies, derived from their best-fit \texttt{Bagpipes} SEDs and assuming the photometric redshifts from \eazy. Overplotted as gray points are measurements from other similarly bright $z>7$ sources in the literature \citep{rb16,wilkins16,hoag17,hashimoto19} for comparison, where available. Black squares and error bars indicate the median and 1$\sigma$ observed scatter in bins of absolute magnitude (left panel; for the $z\sim8$ sample only) and photometric redshift (right panel). The gray filled region in the left panel indicates an approximate range where sources are not expected to be found, based on our galaxy simulations from Section \ref{subsubsec:prior}.}
 \label{fig:beta}
\end{figure*}

We begin by exploring the rest-frame \textit{UV}-continuum slopes ($\beta$, defined such that $f_{\lambda}\propto\lambda^{\beta}$) of our fiducial sample of galaxies. The \textit{UV}-continuum is strongly affected by the age, metallicity, and dust contents of the underlying stellar populations and surrounding gas, thus susceptible to significant uncertainties in the measurements of \textit{UV}-derived quantites (e.g., SFRs). $\beta$ measurements therefore serve as valuable diagnostic for quantifying such quantities in the early Universe \citep{meurer99}, particularly while rest-frame optical diagnostics are unobtainable. Thanks to the number of IR filters on \hst/WFC3 (and \spitzer/IRAC) and their wavelength coverage, measurements of \textit{UV} slopes have become routinely possible for star-forming galaxies at $z=7-10$ (e.g., \citealt{bunker10,bouwens12,dunlop12,finkelstein12,bouwens14a,wilkins16}). Two primary approaches exist to measure this quantity: the first approach is to derive $\beta$ from a fit to all available rest-frame \textit{UV} photometry (a minimum of two bandpasses are required), while the second approach is to measure the slope directly from the best-fit SED over a fixed (rest-frame) wavelength range. In this study we opt for the second approach, since, given the high redshift ranges that our sources probe many sources may not have the two adjacent filters required to make appropriate measurements, or even two filters that are uncontaminated by the Lyman-break and/or possible Lyman-$\alpha$ emission. Additionally, the large range of redshifts utilized in this study requires that $\beta$ measurements make use of different filters according to the redshift bin concerned, to ensure sampling of similar parts of the rest-frame spectrum. Naturally, such concerns introduce uncertainties and make a consistent comparison across redshifts extremely challenging. Thus, here we derive $\beta$ values directly from the best-fit SEDs, using a bootstrap approach and the flux density at a rest-frame wavelength range of 1300-2100 \AA. For each source in our sample, we sample 1000 best-fit SEDs from the posterior distributions of the free parameters and measure the resulting $\beta$ values using a simple power law. We then take the median and standard deviation over the resulting distribution of $\beta$ values as the fiducial value and $1\sigma$ uncertainty. Additionally, we also simultaneously measure the intrinsic \textit{UV} magnitude of each source, $M_{\rm UV}$, assuming the flux density of each SED at a rest-frame wavelength of 1600 \AA\ and a distance modulus. The fiducial value and uncertainty are then derived in identical fashion to the $\beta$ values. We present both sets of galaxy measurements in Figure~\ref{fig:beta} and Table~\ref{tab:uvprops}.

From the left panel of the figure, we find the majority of our points fall under expected and typical values of $\beta$ - i.e., around $-2.5<\beta<-1.5$ - which have been observed in other samples of similarly bright galaxies at high redshift (e.g., \citealt{bouwens14a,wilkins16,hoag17}). Compared to galaxy populations at lower redshift, such especially blue slopes possibly indicate the presence of low metallicity systems with low dust contents, as expected in the early Universe where short dynamical lifetimes do not allow for copious amounts of metal and dust production from aged stars \citet{bouwens12}. For comparison, we also highlight the the $\beta-M_{\rm UV}$ parameter space limits within which galaxies are expected to reside by adopting the simulated galaxy dropout catalogs presented in Section \ref{subsubsec:prior} and running them through identical photo-$z$ selection criteria as the real data. We plot as a shaded region in Figure Figure~\ref{fig:beta} the parameter space onto which the simulated galaxies \textit{do not} fall.

\begin{table*}
\centering
\begin{tabular}{lccc}
\hline
\hline
Field\_ID & $M_{\rm UV}$ & $\beta$ & $r_{50}$ \\
 & [mag] &  & [kpc] \\
\hline
\\
\multicolumn{4}{c}{$z\sim8$ high-confidence candidates} \\
\hline
0314$-$6712\_383 & -23.10$\pm$0.01 & -2.49$\pm$0.05 & $<$0.85 \\
0409$-$5317\_313 & -20.90$\pm$0.06 & -2.37$\pm$0.27 & $<$0.84 \\
0440$-$5244\_742 & -21.37$\pm$0.12 & -2.14$\pm$0.26 & 0.45$\pm$0.01 \\
0830$+$6555\_244 & -21.60$\pm$0.07 & -2.00$\pm$0.23 & 0.79$\pm$0.02 \\
0853$+$0310\_112 & -22.18$\pm$0.03 & -2.45$\pm$0.15 & 0.25$\pm$0.00 \\
0925$+$1360\_899 & -21.38$\pm$0.10 & -2.06$\pm$0.25 & 1.34$\pm$0.04 \\
0948$+$5757\_697 & -21.33$\pm$0.08 & -2.06$\pm$0.25 & 1.05$\pm$0.03 \\
0955$+$4528\_914 & -21.18$\pm$0.09 & -2.09$\pm$0.27 & 0.43$\pm$0.02 \\
0956$+$2848\_986 & -20.53$\pm$0.10 & -1.64$\pm$0.20 & 0.31$\pm$0.01 \\
1017$-$2052\_310 & -21.10$\pm$0.13 & -2.06$\pm$0.29 & 0.22$\pm$0.01 \\
1033$+$5051\_164 & -21.28$\pm$0.07 & -2.12$\pm$0.24 & 0.25$\pm$0.01 \\
1104$+$2813\_447 & -21.37$\pm$0.11 & -2.31$\pm$0.23 & 1.76$\pm$0.06 \\
1218$+$3008\_638 & -21.17$\pm$0.08 & -2.28$\pm$0.26 & 1.14$\pm$0.03 \\
1437$+$5043\_1241 & -21.41$\pm$0.03 & -1.85$\pm$0.17 & 0.44$\pm$0.01 \\
1515$-$1517\_698 & -21.20$\pm$0.09 & -2.25$\pm$0.28 & 0.58$\pm$0.02 \\
1558$+$0812\_601 & -22.05$\pm$0.06 & -2.26$\pm$0.24 & 1.33$\pm$0.04 \\
1917$-$3335\_929 & -21.06$\pm$0.13 & -2.15$\pm$0.29 & 0.73$\pm$0.02 \\
2203$+$1851\_1071 & -21.58$\pm$0.08 & -2.07$\pm$0.21 & 1.74$\pm$0.06 \\
0728$+$0509\_232 & -22.14$\pm$0.04 & -2.03$\pm$0.21 & $<$0.82 \\
0104$+$0021\_339 & -21.75$\pm$0.07 & -2.44$\pm$0.20 & 1.20$\pm$0.02 \\
1149$+$2202\_169 & -21.47$\pm$0.09 & -2.40$\pm$0.23 & 0.79$\pm$0.02 \\
1149$+$2202\_343 & -22.13$\pm$0.08 & -2.13$\pm$0.25 & 0.89$\pm$0.02 \\
2134$-$0708\_2928 & -21.18$\pm$0.11 & -2.29$\pm$0.26 & 1.06$\pm$0.03 \\
0859$+$4114\_138 & -21.84$\pm$0.07 & -2.28$\pm$0.21 & 0.86$\pm$0.02 \\
0859$+$4114\_718 & -21.93$\pm$0.07 & -2.14$\pm$0.23 & 0.64$\pm$0.02 \\
1115$+$2548\_455 & -21.26$\pm$0.08 & -1.94$\pm$0.25 & $<$0.81 \\
\hline
\\
\multicolumn{4}{c}{$z\sim9$ high-confidence candidates} \\
\hline
1607$+$1332\_996 & -21.87$\pm$0.09 & -1.72$\pm$0.15 & 1.20$\pm$0.04 \\
0037$-$3337\_563 & -21.41$\pm$0.09 & -2.21$\pm$0.28 & 0.49$\pm$0.02 \\
1420$+$3743\_1025 & -21.46$\pm$0.10 & -2.18$\pm$0.29 & 0.81$\pm$0.03 \\
\hline
\\
\multicolumn{4}{c}{$z\gtrsim10$ high-confidence candidates} \\
\hline
1459$+$7146\_344 & -21.59$\pm$0.13 & -2.25$\pm$0.29 & $<$0.72 \\
1142$+$3020\_67 & -22.36$\pm$0.13 & -2.31$\pm$0.26 & 1.47$\pm$0.04 \\
\hline
\end{tabular}
\caption{The rest-frame \textit{UV} properties ($M_{\rm UV}$, $\beta$ and half-light radius) for the high redshift galaxies listed in Table~\ref{tab:candidates}.}
\label{tab:uvprops}
\end{table*}

In order to quantify the observed data points, we measured over the full $\beta-M_{\rm UV}$ plane a median slope of $\beta=-2.15$ with a standard deviation of $0.2$. We do not observe any obvious or significant trends between the samples, although there is some tentative evidence to suggest more luminous galaxies display bluer slopes. At face-value, such an interpretation would prove misleading, since this may imply that more massive galaxies represent the most dust-free and metal-poor systems, an interpretation which is inconsistent with the majority of other findings (e.g., \citealt{bouwens12}). Given that those data points also follow a redshift trend, a likely explanation is that we are simply observing the most massive galaxies at each epoch, which in turn appear more dust-free and metal-poor at higher redshifts. Such an interpretation is supported by the panel on the right of Figure~\ref{fig:beta} (discussed below). To better quantify whether any trend exists, we derive median values in bins of $\beta$ and $M_{\rm UV}$ and plot these as black squares. To ensure the points are not impacted by redshift trends and to minimize small number effects as much as possible, we apply such measurements to our $z\sim8$ points only. The median data further illustrate the lack of any real trend over our dynamic $M_{\rm UV}$ range (approximately 2 magnitudes). The binned quantities are presented in (the top half of) Table~\ref{tab:slopes}.

While small sample sizes prevent us from inspecting the trend as a function of absolute magnitude \textit{and} redshift, we plot on the right panel of Figure~\ref{fig:beta} the \textit{UV} slopes a function of photometric redshift. Again, no obvious trend is found between the two quantities. As with the $\beta-M_{\rm UV}$ plane, we again plot median values of $\beta$ and photometric redshift in order to better highlight potential trends and present the resulting values in (the lower half of) Table~\ref{tab:slopes}. Once again no clear trend is found.


\begin{table*}
\centering
\begin{tabular}{lccccccc}
\hline
\hline
$M_{\rm UV}$ bin & $M_{\rm UV,mean}$ & $M_{\rm UV,median}$ & $\sigma_{M_{\rm UV}}$ & $\beta_{\rm mean}$ & $\beta_{\rm median}$ & $\sigma_{\beta}$ & $N_{\rm gals}$ \\
\hline
-20.50 to -21.00 & -20.71 & -20.71 & 0.19 & -2.00 & -2.00 & 0.36 & 2 \\
-21.00 to -21.50 & -21.27 & -21.27 & 0.12 & -2.14 & -2.13 & 0.15 & 14 \\
-21.50 to -22.00 & -21.74 & -21.75 & 0.14 & -2.19 & -2.14 & 0.16 & 5 \\
-22.00 to -22.50 & -22.13 & -22.14 & 0.05 & -2.22 & -2.19 & 0.16 & 4 \\
\hline
\\
\hline
\hline
$z_{\rm phot}$ bin & $z_{\rm phot,mean}$ & $z_{\rm phot,median}$ & $\sigma_{z_{\rm phot}}$ & $\beta_{\rm mean}$ & $\beta_{\rm median}$ & 1$\sigma_{\beta}$ & $N_{\rm gals}$ \\
\hline
7.50 to 8.50 & 7.86 & 7.75 & 0.25 & -2.17 & -2.14 & 0.19 & 26 \\
8.50 to 9.50 & 8.76 & 8.76 & -- & -2.04 & -2.18 & 0.22 & 3 \\
9.50 to 10.50 & 9.89 & 9.89 & 0.11 & -2.28 & -2.28 & 0.03 & 2 \\
\hline
\end{tabular}
\caption{The mean and median values and 1$\sigma$ (observed) scatter of the \textit{UV}-continuum slopes measured over our fiducial sample of galaxies (detections only), as a function of absolute \textit{UV} magnitude (top) and photometric redshift (bottom). NB: the top half of the table makes use of the  $z\sim8$ data points only.}
\label{tab:slopes}
\end{table*}


\subsection{The sizes of galaxies}
\begin{figure*}
\center
 \includegraphics[width=0.8\textwidth]{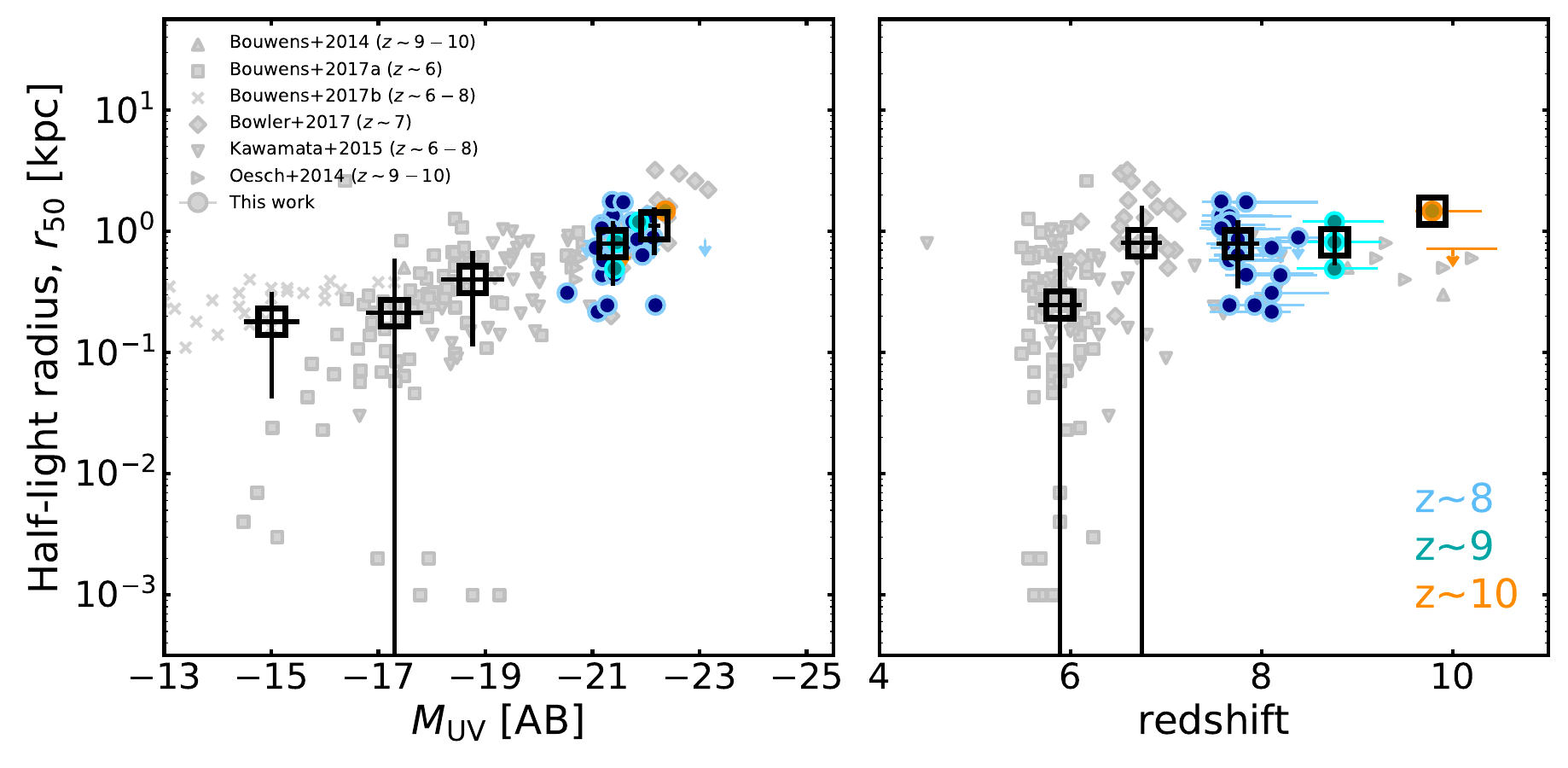}
 \caption{The galaxy half-light radius estimated with \texttt{SExtractor} on the detection images of our galaxies, versus $M_{\rm UV}$ (left) and photometric redshift (right). The uncertainties on the radius derive from the uncertainty on the angular diameter distance due to the lower and upper photometric redshift bounds from \eazy.}
 \label{fig:sizemag}
\end{figure*}

We next inspect the intrinsic sizes of our high-$z$ candidates, since these reveal clues as to the assembly history of galaxies and are often invoked as part of selection criteria for e.g., luminous $z>9$ sources, where reliable photometric constraints from \spitzer/IRAC are often lacking to distinguish with lower redshift interlopers (see discussions in \citealt{holwerda15,holwerda20}). Here we adopt the rest-frame \textit{UV} half-light radii ($r_{\rm 50}$) of our samples using the \texttt{SExtractor} (non-parametric) half-light radius, $r_{\rm 50, SExtractor}$, measured on the detection images. Each measured radius is converted to arcsecond units and corrected for the filter PSF, $r_{\rm PSF}$, such that $r_{\rm 50}=\sqrt{r_{\rm 50, SExtractor}^{2}-r_{\rm PSF}^{2}}$. While the F160W PSF is measured at $r_{\rm PSF}=0\farcs14$, \citet{morishita21} showed the measurement limit for galaxy sizes is closer to $r=0\farcs17$. Thus, we assume the latter quantity as the lower limit to measured \texttt{SExtractor} sizes. Each $r_{\rm 50}$ value is then converted to a physical size based on its angular size and redshift and presented along with $M_{\rm UV}$ magnitudes and the photometric redshifts of the galaxies in Figure~\ref{fig:sizemag} and Table~\ref{tab:uvprops}.

Given the relatively limited range in magnitude of our sample, the expected slope of the size-luminosity relation is marginally significant (see binned data in the Figure). By enlarging the range of luminosities (by $\sim6$ mag down to $M_{\rm UV}\sim-14$ mag) with available data from lensing clusters and blank fields (e.g., \citealt{kawamata15,bouwens14a,oesch14,bowler17,bouwens17a,bouwens17b}), the combined samples clearly reveal an obvious size-mass relationship, with the SuperBoRG sample nicely populating the high luminosity end. The most massive and luminous galaxies display the largest sizes (i.e., a few kpc) and the faintest the smallest sizes (sub-kpc down to several pc, although such measurements are uncertain given the resolution limits of \textit{Hubble} and the uncertainties of lensing models typically adopted to search for the faintest sources - c.f. \citealt{bouwens17a}, \citealt{bouwens17b} and \citealt{yang21}). As an illustration, the median size and 1$\sigma$ scatter found at the brightest end of the SuperBoRG sample is 0.80$\pm$0.44 while this decreases to 0.18$\pm$0.14 for faint ($M_{\rm UV}\sim-15$ mag) galaxies, nearly a factor of $4.5\times$ smaller. Such a relation has been observed at both low and high redshift, with lower redshift results also highlighting the well-established redshift evolution of the relation (e.g., \citealt{vanderwel14,yang21}). However, it remains unclear whether such an evolution extends to redshifts of $z\gtrsim7$, where evolutionary times decrease dramatically.

Analytical models have predicted a scaling of galaxy sizes with redshift of (1+$z$)$^{b}$, with values of $b$ ranging between -1.5 and -1 \citep{fall80,mo98}, however the census from observational data sets remains unclear and is dominated by scatter and small sample sizes. For instance, making use of bright galaxy samples found over the UltraVISTA/COSMOS and UDS/SXDS fields, \citet{bowler17} found several $z\sim7$ sources showing evidence of extended shapes and merging systems, however studies making use of smaller samples of $7\lesssim z\lesssim9$ galaxies from the \textit{Hubble} Ultra Deep Field, eXtremely Deep Field, or CANDELS fields, appear to find more compact objects at higher redshift (e.g., \citealt{oesch14,holwerda15}). Making use once again of the enlarged sample size from SuperBoRG and results from the literature, we find no clear trend of galaxy size with redshift (over individual points or median values). In particular, focusing on the SuperBoRG data only - which greatly extends the sample sizes of bright, $z\gtrsim8$ galaxies - we find evidence for both compact ($\sim$0.1 kpc) and more extended ($\sim$1 kpc) sources.


\begin{table*}
\centering
\begin{tabular}{lccccccc}
\hline
\hline
$M_{\rm UV}$ bin & $M_{\rm UV,mean}$ & $M_{\rm UV,median}$ & $\sigma_{M_{\rm UV}}$ & $r_{\rm 50,mean}$ & $r_{\rm 50,median}$ & $\sigma_{r_{\rm 50}}$ & $N_{\rm gals}$ \\
\hline
-14.00 to -16.00$^{\dagger}$ & -15.10 & -15.01 & 0.52 & 0.17 & 0.18 & 0.14 & 24 \\
-16.00 to -18.00$^{\dagger}$ & -17.19 & -17.30 & 0.53 & 0.26 & 0.21 & 0.39 & 45 \\
-18.00 to -20.00$^{\dagger}$ & -18.92 & -18.76 & 0.59 & 0.44 & 0.40 & 0.29 & 55 \\
-20.00 to -22.00 & -21.38 & -21.38 & 0.31 & 0.83 & 0.79 & 0.43 & 22 \\
-22.00 to -24.00 & -22.18 & -22.15 & 0.11 & 0.98 & 1.11 & 0.48 & 4 \\
\hline
\\
\hline
\hline
$z_{\rm phot}$ bin & $z_{\rm phot,mean}$ & $z_{\rm phot,median}$ & $\sigma_{z_{\rm phot}}$ &  $r_{\rm 50,mean}$ & $r_{\rm 50,median}$ & $\sigma_{r_{\rm 50}}$ & $N_{\rm gals}$ \\
\hline
5.50 to 6.50$^{\dagger}$ & 5.91 & 5.89 & 0.23 & 0.34 & 0.25 & 0.37 & 97 \\
6.50 to 7.50$^{\dagger}$ & 6.79 & 6.75 & 0.22 & 1.14 & 0.80 & 0.83 & 26 \\
7.50 to 8.50 & 7.81 & 7.75 & 0.22 & 0.83 & 0.79 & 0.45 & 22 \\
8.50 to 9.50 & 8.76 & 8.76 & -- & 0.84 & 0.81 & 0.29 & 3 \\
9.50 to 10.50 & 9.79 & 9.79 & -- & 1.47 & 1.47 & -- & 1 \\
\hline
\end{tabular}
\caption{The mean and median values and 1$\sigma$ (observed) scatter of the half-light radius, $r_{\rm 50}$, measured over our fiducial sample of galaxies (detections only), as a function of absolute \textit{UV} magnitude (top) and photometric redshift (bottom). Values brighter than $M_{\rm UV}=-20$ mag make use of the SuperBoRG data only, while values fainter than that limit make use of the data in the literature highlighted in Figure~\ref{fig:sizemag} (bins marked with $\dagger$). The same applied for redshifts greater or less than $z=7.5$. As in Table~\ref{tab:slopes}, the 1$\sigma$ column refers to the scatter of data points in each bin.}
\label{tab:radii}
\end{table*}

\subsection{Rest-frame optical properties}
\label{sec:opticalprops}
\begin{figure*}
\center
 \includegraphics[width=0.9\textwidth]{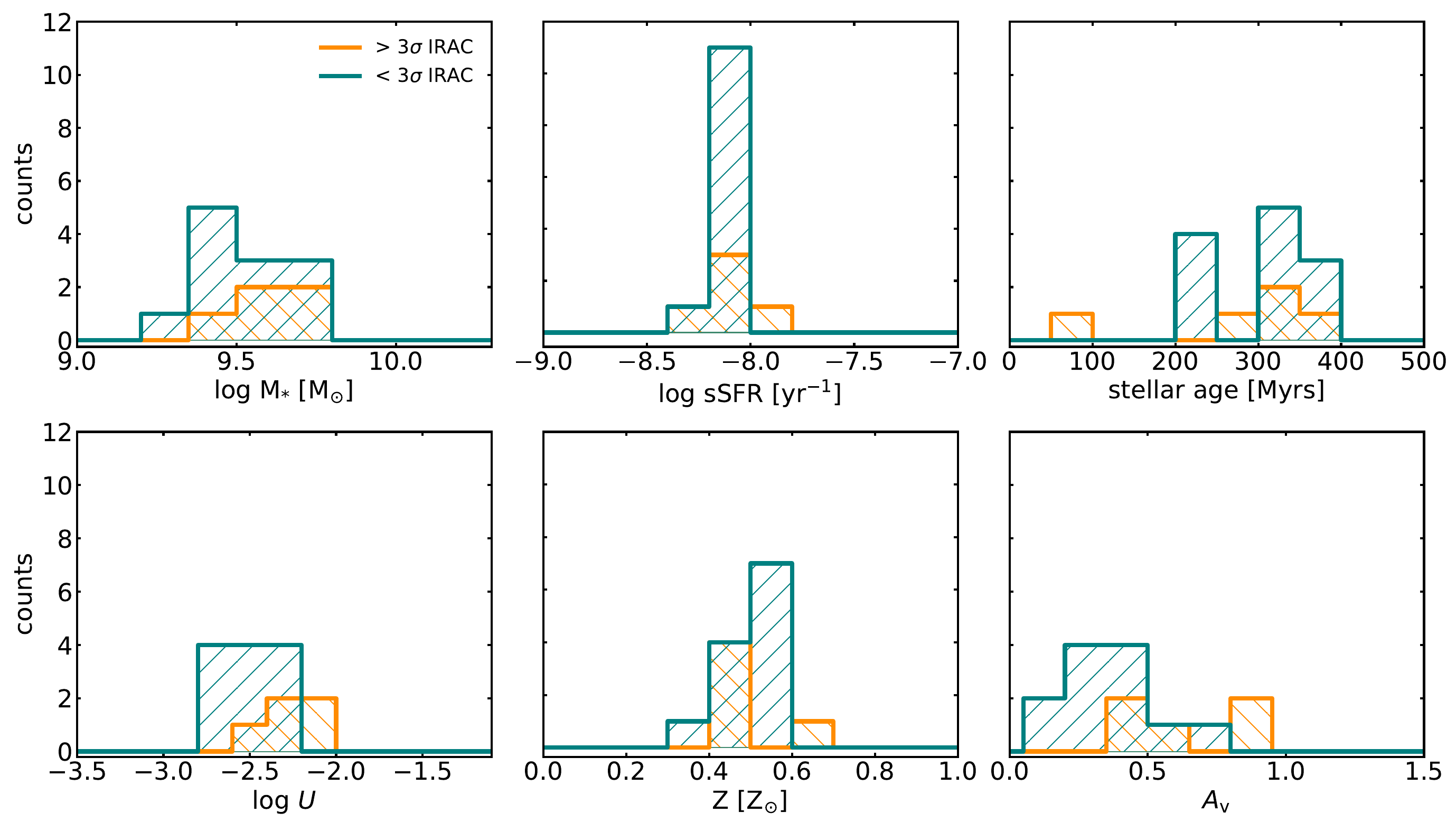}
 \caption{The rest-frame optical global galaxy properties (from left to right, top to bottom: stellar mass, specific SFR, stellar age, ionization parameter, metallicity and dust) of SuperBoRG galaxies with $>3\sigma$ (orange histogram) and $<3\sigma$ (teal histogram) \spitzer/IRAC detections, derived using \texttt{Bagpipes} SED fitting on the available \hst+\spitzer photometry. Quantities for individual objects are tabulated in Table~\ref{tab:opticalprops}.}
 \label{fig:optprops}
\end{figure*}

For galaxy candidates where \spitzer/IRAC coverage is available, we analyze the best-fit physical parameters from \texttt{Bagpipes}, namely the stellar masses, star formation rates and stellar ages. Since these quantities are primarily derived from the rest-frame optical at $\gtrsim$3 $\mu$m, the \spitzer/IRAC 3.6 $\mu$m and 4.5 $\mu$m band constraints are crucial. 18 galaxies from our fiducial sample have \spitzer\ coverage in the 3.6 $\mu$m and/or the 4.5 $\mu$m band. Many of these are only upper limits, however even upper limits can yield important constraints. We thus separate out galaxies with high confidence ($>3\sigma$) IRAC detections in one or both bands (5 sources) and those without (12 sources). For both samples we illustrate the results as histograms in Figure~\ref{fig:optprops} and tabulate the results in Table~\ref{tab:opticalprops}.

We begin by noting that our sample display a small range of values in their global properties. We find stellar masses ranging from log\,M$_{*}$/M$_{\odot}\simeq$9.25-9.75, specific SFRs from $-8.5\lesssim$log\,sSFR$\lesssim-7.75$, stellar ages of a few hundreds of Myrs, ionization parameters of $-3.0\lesssim$log\,$U\lesssim-2.0$, metallicities of approximately 0.3-0.7 solar and a range of dust obscuration from $0\lesssim A_{\rm v}\lesssim$1 mag. Such values are consistent with and remarkably similar to others inferred through photometry for similarly bright, $z\sim8$ objects selected from lensing clusters (e.g., RELICS; \citealt{strait20}) or legacy and blank fields such as CANDELS (e.g., \citealt{labbe13,stark17}) and BoRG (e.g., \citealt{bridge19}), pointing to uniformity in the properties of the most massive Lyman Break selected galaxies and potentially little evolution in their rest-frame optical properties, at least between $8\lesssim z\lesssim10$.

However, while the estimated sSFR is not particularly sensitive to the assumed SFH, recent studies have shown the contamination of the \spitzer/IRAC bands by nebular emission lines (in particular [\OIII]+H$\beta$) at redshifts of $z\simeq7-9$ \citep{labbe13,rb16}, making precise continuum or line emission measurements particularly challenging. For instance, for a sample of especially bright ($M_{\rm UV}\sim-21$ mag) galaxies assumed to be intense (EW$>1000$ \AA) line emitters, \citet{rb20} found that reducing the line emission to more representative strengths could increase the stellar mass of the galaxies by a factor of $\sim30\times$. This is particularly relevant for the bright, $z>8$ galaxies we examine here, however more precise measurements are currently beyond the capabilities of current telescopes and will require the sensitive spectro-photometric capabilities of the \textit{James Webb} Space Telescope (see e.g., \citealt{rb21}).

\begin{table*}
\centering
\begin{tabular}{lcccccc}
\hline
\hline
Field\_ID & log M$_{*}$ & log sSFR & $A_{\rm v}$ & log $U$ & Z & Stellar age \\
& [M$_{\odot}$] & [yr$^{-1}$] & [mag] &  & [Z$_{\odot}$] & [Myrs] \\
\hline
\multicolumn{7}{c}{$>3\sigma$ IRAC selection} \\
\hline
0925$+$1360\_899 & 9.62$\pm$0.25 & -8.16$\pm$0.15 & 0.50$\pm$0.22 & -2.44$\pm$0.85 & 0.46$\pm$0.26 & 327.67$\pm$173.80 \\
0956$+$2848\_986 & 9.69$\pm$0.22 & -8.21$\pm$0.12 & 0.85$\pm$0.15 & -2.06$\pm$0.84 & 0.50$\pm$0.24 & 381.36$\pm$135.13 \\
1033$+$5051\_164 & 9.47$\pm$0.30 & -8.13$\pm$0.15 & 0.42$\pm$0.22 & -2.10$\pm$0.84 & 0.44$\pm$0.28 & 299.67$\pm$173.42 \\
2203$+$1851\_1071 & 9.65$\pm$0.22 & -8.14$\pm$0.16 & 0.49$\pm$0.20 & -2.28$\pm$0.81 & 0.49$\pm$0.28 & 302.60$\pm$171.40 \\
1607$+$1332\_996 & 9.77$\pm$0.17 & -7.93$\pm$0.06 & 0.88$\pm$0.15 & -2.32$\pm$0.86 & 0.60$\pm$0.24 & 88.99$\pm$78.19 \\
\hline
\multicolumn{7}{c}{$<3\sigma$ IRAC selection} \\
\hline
0830$+$6555\_244 & 9.68$\pm$0.20 & -8.06$\pm$0.14 & 0.56$\pm$0.23 & -2.62$\pm$0.86 & 0.56$\pm$0.26 & 231.77$\pm$155.28 \\
0853$+$0310\_112 & 9.59$\pm$0.22 & -8.21$\pm$0.15 & 0.11$\pm$0.10 & -2.42$\pm$0.80 & 0.42$\pm$0.27 & 382.92$\pm$170.13 \\
0948$+$5757\_697 & 9.58$\pm$0.26 & -8.16$\pm$0.15 & 0.47$\pm$0.22 & -2.37$\pm$0.83 & 0.58$\pm$0.26 & 323.95$\pm$172.06 \\
1017$-$2052\_310 & 9.45$\pm$0.34 & -8.14$\pm$0.15 & 0.45$\pm$0.25 & -2.53$\pm$0.82 & 0.52$\pm$0.27 & 311.69$\pm$160.04 \\
1104$+$2813\_447 & 9.37$\pm$0.29 & -8.17$\pm$0.16 & 0.27$\pm$0.20 & -2.60$\pm$0.80 & 0.45$\pm$0.28 & 338.79$\pm$186.51 \\
1218$+$3008\_638 & 9.37$\pm$0.30 & -8.19$\pm$0.16 & 0.31$\pm$0.22 & -2.64$\pm$0.82 & 0.48$\pm$0.29 & 362.25$\pm$183.06 \\
1437$+$5043\_1241 & 9.73$\pm$0.17 & -8.05$\pm$0.14 & 0.70$\pm$0.19 & -2.26$\pm$0.81 & 0.55$\pm$0.25 & 222.25$\pm$142.16 \\
0104$+$0021\_339 & 9.43$\pm$0.24 & -8.19$\pm$0.16 & 0.16$\pm$0.15 & -2.63$\pm$0.83 & 0.35$\pm$0.30 & 361.28$\pm$179.48 \\
2134$-$0708\_2928 & 9.31$\pm$0.30 & -8.17$\pm$0.16 & 0.26$\pm$0.22 & -2.56$\pm$0.83 & 0.44$\pm$0.29 & 341.57$\pm$190.62 \\
0859$+$4114\_138 & 9.56$\pm$0.27 & -8.14$\pm$0.15 & 0.29$\pm$0.19 & -2.37$\pm$0.85 & 0.51$\pm$0.27 & 311.68$\pm$173.93 \\
0859$+$4114\_718 & 9.65$\pm$0.21 & -8.03$\pm$0.15 & 0.43$\pm$0.24 & -2.28$\pm$0.82 & 0.55$\pm$0.27 & 209.42$\pm$163.33 \\
1459$+$7146\_344 & 9.46$\pm$0.29 & -8.06$\pm$0.12 & 0.38$\pm$0.25 & -2.47$\pm$0.86 & 0.52$\pm$0.29 & 238.50$\pm$128.50 \\
\hline
\end{tabular}
\caption{The global galaxy properties and 1$\sigma$ posterior distribution errors derived from realistic SFHs from \texttt{Bagpipes}, for galaxies where \spitzer/IRAC coverage is available. The fits make use of the ``high-$z$'' parameter ranges tabulated in Table~\ref{tab:params} and the values quoted here indicate the median and 1$\sigma$ from the resulting posterior distribution of each quantity. The samples are separated into galaxies with $>3\sigma$ IRAC detections (top) and those with $<3\sigma$ detections (bottom).}
\label{tab:opticalprops}
\end{table*}

\section{Keck/MOSFIRE Observations}
\label{sec:keck}
In order to confirm a subset of our selected sources, we endeavoured to obtain deep near-IR spectroscopy of two $z\sim8$ galaxies from our present sample with Keck/MOSFIRE, and present previously unpublished follow up spectroscopy of two other BoRG galaxies, with the aim of detecting \lya\ emission in $Y-$band. The two sources selected for follow up were 0853$+$0310\_112 and 2203$+$1851\_1071, with apparent magnitudes of $H_{\rm 160}\sim25-25.5$ AB and photometric redshifts of $z_{\rm phot}=7.66$ and $z_{\rm phot}=7.84$, respectively. As mentioned in Section \ref{sec:compare}, the observations of 0853$+$0310\_112 have already been discussed in \citet{morishita20} and we refer to that paper for details. As a brief summary, however, $\sim4$ hrs of $Y$-band did not reveal any \lya\ emission. Here we present the follow up observations of the second object, 2203$+$1851\_1071. The data were obtained over two nights (25th November 2020 and 7th December 2020; UC2020B, PI T. Treu) in $Y-$band spectroscopic mode, adopting an ABBA dither pattern with 1.25$''$ offsets. Each exposure was set to 180 $s$ at a spectral resolution of $R=3500$ (for a slit 0.7$''$ wide) and the weather was mostly clear ($<1''$). Each mask contained at least one star with a slitlet placed on top in order to measure the sky conditions and transparency. Discarding all exposures with a seeing of $>1''$, the resulting data were reduced with the standard MOSFIRE data reduction pipeline (DRP) and visually inspected for any resulting emission lines. No clear emission line was found.

\begin{figure}
\center
 \includegraphics[width=\columnwidth]{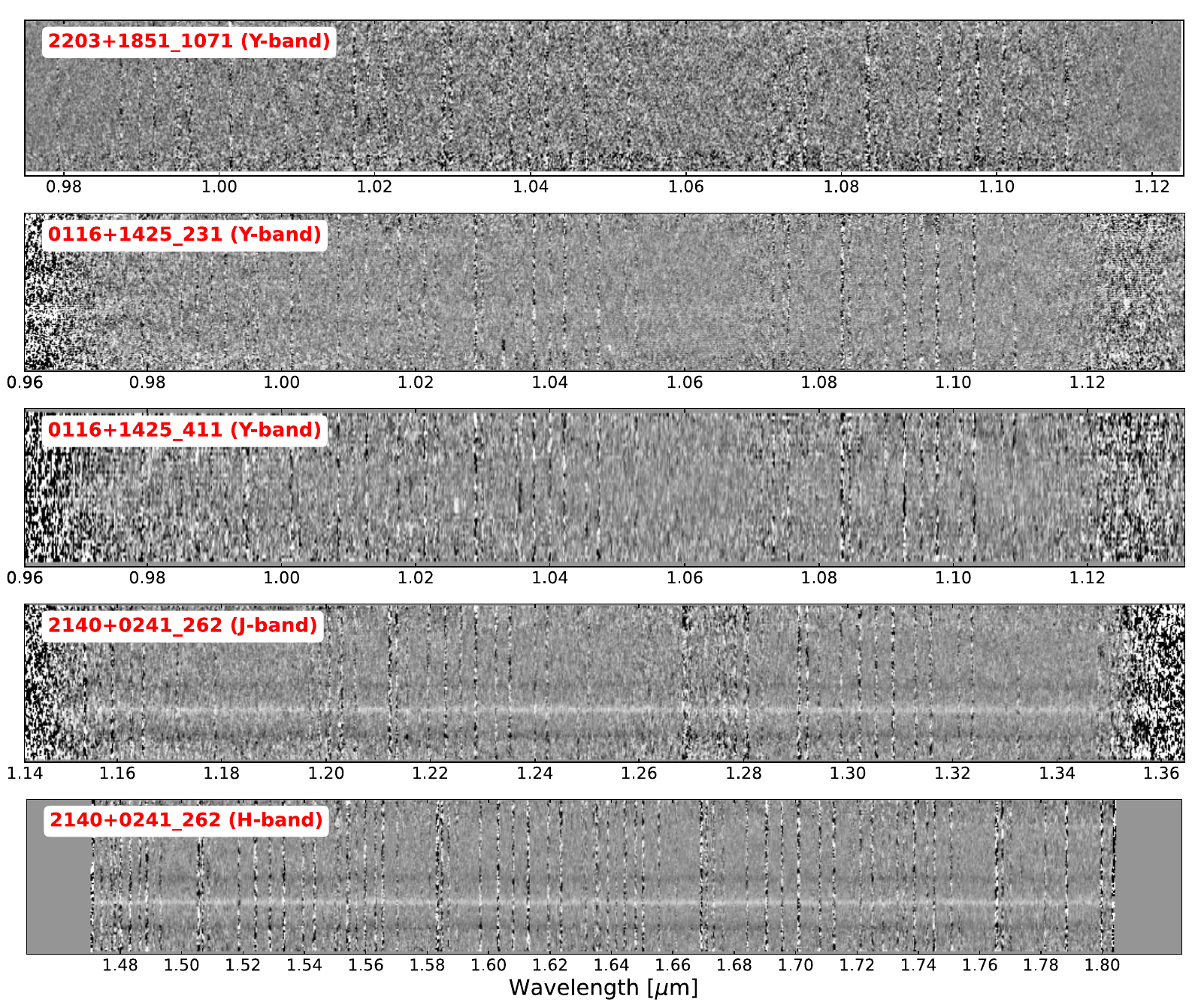}
 \caption{The 2D Keck/MOSFIRE spectra (\textit{Y-}, \textit{J-} and \textit{H-}band) of one of our selected $z\sim8$ galaxies (2203$+$1851\_1071), as well as archival data over previously-selected BoRG targets (two $z\sim8$ and one $z\sim10$ candidate) not selected in this study (0116$+$1425\_231, 0116$+$1425\_411 and 2140$+$0241\_262). The long trace found in the 2140$+$0241\_262 spectra are of a low-$z$ galaxy that was present in the MOSFIRE slit and do not correspond to the high-$z$ candidate.}
 \label{fig:spectra}
\end{figure}

Finally, we also present previously-unpublished archival Keck/MOSFIRE observations of two additional BoRG fields with 3 targets selected as part of previous studies (but not selected here), namely 2140$+$0241 (1 candidate) and 0116$+$1425 (2 candidates). The 2140$+$0241 field was observed on August 21 2019 for $\sim2$\,hrs with $J$-band and $\sim1$\,hr with $H$-band. During the observations, the seeing was stable ($\sim0.\!\arcsec6$). The data were reduced with DRP as well as a pipeline developed by the MOSDEF team \citep{kriek15}. No emission lines were identified along the trace of a $z\sim10$ candidate identified in \citet[][]{morishita18}. The 0116$+$1425 field, where two $z\sim8$ candidates reported by \citet[][also \citealt{livermore18}]{calvi16} are located, was observed on the same night and September 19 2019 for 1.8\,hrs and 1\,hr, respectively, with the $Y$-band. The seeing was moderate ($\sim0.\!\arcsec8$) and poor ($\sim1\arcsec$) on those nights. The second night was partially cancelled due to a technical issue. The data were reduced in the same way as the 2140$+$0241 data and no emission lines were detected. All reduced 2D spectra are displayed in Figure \ref{fig:spectra}.

The non-detection of clear \lya\ in $>3$ hrs of MOSFIRE observations for each of the 2 targets selected in this study is suggestive of two main points, namely (i) that our selection procedures are robust against low-$z$ ($1\gtrsim z\gtrsim3$) interlopers, which would reveal clear emission line fluxes in only a few hours' worth of exposure time (e.g., \citealt{kriek15}), and (ii) that the attenuation of \lya\ is primarily due to the resonant scattering of the line by an opaque IGM at $z\sim8$. This latter consideration is consistent with the picture presented by our previous spectroscopic campaigns of a highly neutral IGM at $z\sim8$. In \citet{treu13}, we conducted MOSFIRE spectroscopic follow up of 13 candidate $z\sim8$ BoRG galaxies, of which none revealed any clear \lya\ emission in 1-3 hrs of observations. Combining the non-detections with other results from the literature, the analysis found that the fraction of \lya\ emitters amongst samples of Lyman-break galaxies decreased by a factor of $\sim3\times$ from $z\sim6$ to $z\sim8$ (only $\sim300$ Myrs of evolutionary time), attributable to an increasingly opaque neutral medium. While some recent results of especially luminous $z\sim8$ galaxies (e.g., \citealt{finkelstein15,oesch15,zitrin15,rb16,laporte17,stark17,hashimoto18,laporte21}) have begun to challenge this picture with unexpected detections of \lya\ in objects with a large \spitzer/IRAC 3.6 $\mu$m minus 4.5 $\mu$m excess, such galaxies come from small area surveys (e.g., CANDELS or Frontier Fields clusters) that are prone to cosmic variance (and clustering) effects and whether their extreme properties are representative of the $z\gtrsim8$ general population is subject to debate. The observations presented here and in \citet{treu13} are likely more robust to cosmic variance effects due to their (pure-)parallel selection and are more consistent with an especially opaque IGM at $z\sim8$.

\section{The $z\gtrsim8$ Cosmic SFR Density}
\label{sec:csfrd}
One of the most interesting questions to ask at $z\gtrsim8$ is whether the redshift evolution of the cosmic SFR density (CSFRD) matches that of the dark matter halo mass function, or whether an increase in star formation efficiency is required to match observed trends. However, such determinations rely heavily on an assumed conversion factor to convert \textit{UV} luminosities to star formation rates, which are typically calibrated for low-redshift samples. In addition to being an inherently uncertain procedure, the uncertainties are compounded by the fact that such calibrations are often based on a single tracer of star formation, which can heavily underestimate or bias the true amount of global star formation in a galaxy - for instance, measurements from H$\alpha$ or the \textit{UV} continuum only trace recent (a few 10$^{6}$ yrs) bursts of star formation from massive, young stars and are subject to dust obscuration, while measurements from the far-infrared probe dust-enshrouded star formation and longer-lived stars. It is imperative, therefore, to test whether such assumptions and differences play a major role in derivations of high redshift ($z\gtrsim8$) CSFRD estimates or not. Having constructed a sample of galaxy candidates through NIR color cuts and photo-$z$ estimations that is robust against cosmic variance effects, and derived their global properties, we are in a position to perform such a test using their SED-derived SFRs and compare results with those presented in the literature.

\subsection{Galaxy Completeness Simulations \& Effective Volume}
The computation of a CSFRD requires the (accurate) determination of the effective search volume probed by the data sets in question, which can be written as:

\begin{equation}
V_{\rm eff,i}(z,M_{\rm UV}) = \int P_{i}(z,M_{\rm UV})\frac{dV}{dzd\Omega}d\Omega_{i}dz [\rm Mpc^{3}]
\end{equation}

where $V_{\rm eff,i}$ is the effective volume in Mpc$^{3}$, $P_{i}(z,M_{\rm UV})$ is the ratio of the number of recovered simulated galaxies (i.e., galaxies that are detected in our images and satisfy the selection criteria outlined in Section \ref{sec:sampselect}) at redshift $z$ with \textit{UV} magnitude $M_{\rm UV}$ divided by the number of simulated galaxies at the same redshift with an intrinsic magnitude $M_{\rm UV, int.}$. $i$ refers to the SuperBoRG field in question. 

\begin{figure}
\center
 \includegraphics[width=\columnwidth]{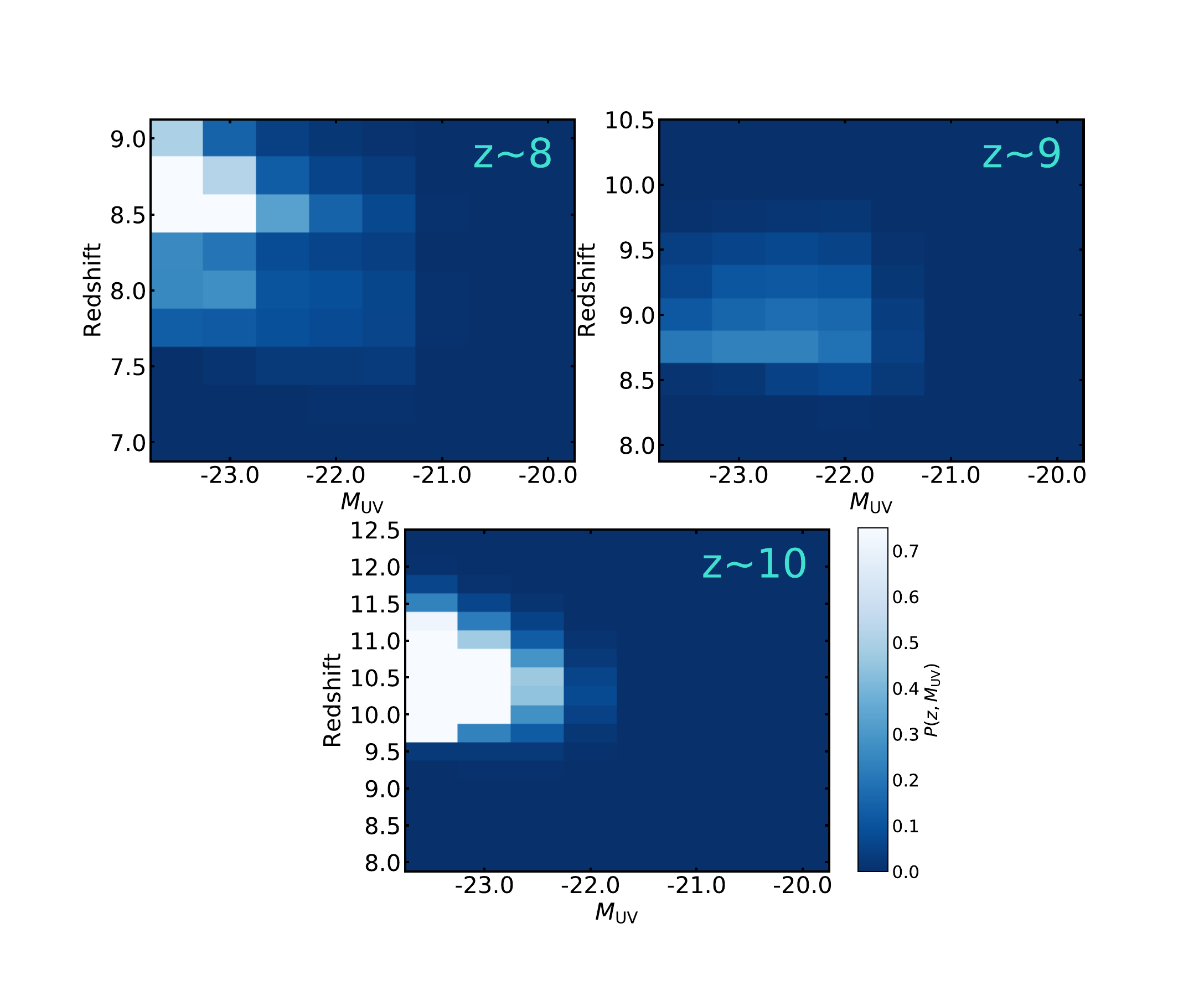}
 \caption{The $P(z,M_{\rm UV})$ values resulting from simulations of high-$z$ galaxy completeness and selection functions for our $z\sim8$ (left), $z\sim9$ (right) and $z\sim10$ (bottom) selection criteria. Identical contrasts are used for each panel.}
 \label{fig:CS}
\end{figure}

The simulations are performed only for sources satisfying all of our $z\sim8$, $z\sim9$ and $z\sim10$ (photo-$z$) selection criteria. More specifically, in order to construct the $P_{i}(z,M_{\rm UV})$ matrix, our simulations consist of inserting artificial high-$z$ sources into our SuperBoRG science images and recovering them in identical fashion to the real ones. In order to construct realistic catalogs of high-$z$ galaxies, we adopt the procedure described in \citet{leethochawalit21} and refer to that paper for a full description. In short, for a given $M_{\rm UV}-z$ bin, 40 different SEDs are extracted from the v1.2 JAdes extraGalactic Ultradeep Artificial Realizations (JAGUAR; \citealt{williams18}) catalogs (adopting the strategy described by \citealt{rb21}) and $M_{\rm UV}$ values from a uniform distribution over the magnitude ranges of interest. The SEDs are subsequently normalized at (rest-frame) 1600 \AA\ by the $M_{\rm UV}$ values and shifted to the required redshift. Each SED is then assigned 200 galaxy postage stamps, constructed based on S\'ersic brightness profiles, each of which is normalized to the required apparent brightness in each relevant passband using the \texttt{GLACiER2}\footnote{\url{https://github.com/nleethochawalit/GLACiAR2-master/}} code and added to the SuperBoRG science image before going through an identical extraction and selection process as the real data. Such a procedure is repeated for each $M_{\rm UV}-z$ bin in a grid of $M_{\rm UV}=\{-23.5,-19.5\}$ mag with $\Delta M_{\rm UV}=0.5$ mag and $z=\{7,9\}$ (for $z\sim8$ dropouts, otherwise $z=\{8,12.5\}$ is used for $z\sim9$ and $z\sim10$ samples) with $\Delta z=0.25$. The proportion of artifical galaxies recovered are then used to determine the redshift- and magnitude-dependent $P(z,M_{\rm UV})$ values, which accounts for completeness. We present the results for $z\sim8$, $z\sim9$ and $z\sim10$ searches in Figure~\ref{fig:CS}. The comoving volume ($v_{\rm eff}$) for each field, $i$, is then computed using the effective search area (converted to a solid angle) and the $P(z,M_{\rm UV})$ values over various magnitude ranges of interest. The total effective search volume ($V_{\rm eff}$) per magnitude bin (see the discussion in the following section) is then taken as the sum of each field's effective volume.

\subsection{Consistency at the Bright End?}
Upon constructing accurate selection volumes, we are now in a position to determine the bright end of the $z\sim8-10$ CSFRD. Previous works have typically made use of a luminosity density or function and a conversion factor to determine a SFRD (e.g., \citealt{bouwens14b,oesch14,oesch18}). However, such an approach is subject to important uncertainties, namely the fact that it is based purely on a number count of \textit{UV}-selected objects and a conversion factor.

Thus, given we have fitted each of our high-$z$ galaxies with detailed SED-fitting and realistic SFHs, we opt instead to use the SFRs produced by \texttt{Bagpipes} (defined as the weighted mean of the SFH over the last 100 Myrs of the galaxy's lifetime), which serve as a crucial baseline for LF measurements. We remind the reader that our SED-fitting accounts for dust effects (typically $A_{\rm v}<0.5$ mag, see Section \ref{sec:opticalprops}), metallicity and age effects, which a LF can only account for statistically. We define the final SFRD in a given magnitude and redshift bin as:

\begin{equation}
\phi(z,M_{\rm UV}) = \frac{\sum \limits_{i=0}^{N} \rm SFR_{i}(\textit{z},M_{\rm UV})}{\sum \limits_{i=0}^{N} V_{\rm eff,i}(M_{\rm UV})} [{\rm M}_{\odot}{\rm yr}^{-1}{\rm Mpc}^{-3}]
\end{equation}

where $i$ again refers to a field of interest. In the calculation we multiply each galaxy SFR by its $P_{\rm gal}$ statistic (i.e., the probability of having a galaxy nature as opposed to a brown dwarf nature) derived in Section \ref{subsubsec:contamination}, and include the 1$\sigma$ uncertainties on the posterior distribution of the SFR. We additionally include a relative shot noise term on the summed value to account for sampling errors from small samples, based on the lower and upper limits presented in \citet{gehrels86}. We conduct our analysis over magnitude bins centered at $M_{\rm UV}=\{-23,-22,-21\}$ mag (with a bin size of $\Delta M_{\rm UV}=1$ mag), where our searches for high-$z$ galaxies yield the largest numbers of candidates. Our lower magnitude limit of $M_{\rm UV}<-20.5$ mag corresponds to a lower SFR limit of 7.9 M$_{\odot}$yr$^{-1}$, representing the brightest end of the luminosity function.

For comparison, we also derive SFRDs from $z\sim8-10$ LFs as found by \citet{schmidt14}, \citet{mason15}, \citet{mcleod16}, \citet{oesch18}, \citet{stefanon19}, \citet{bowler20} and \citet{bouwens21}, integrating each of them between the same magnitude limits as our observed data sets and consistently assuming a $L_{\rm UV}-$SFR$_{\rm UV}$ conversion factor throughout. For the latter consideration, we assume the widely-used factor from \citet{madau14}, namely $\kappa_{UV}=1.15\times10^{-28} M_{\odot}{\rm yr}^{-1}{\rm erg} {\rm s}^{-1} {\rm Hz}^{-1}$. However, given our data points make use of SFRs from \texttt{Bagpipes}, which adopts stellar population synthesis (SPS) models from \citet{bc03} and a \citet{kroupa02} initial mass function (IMF), it is important to adjust the conversion factor prior to application to the data, for consistent comparisons. We thus decrease $\kappa_{UV}$ by 5\% to adjust from GALXEV to \citet{bc03} SPS models and multiply the resulting value by 0.67 to adjust to Kroupa IMF (see Section 3.1.1 of \citealt{madau14}). Furthermore, such a conversion factor does not account for dust obscuration effects, which, while not expected to be significant at $z\geq8$ could nonetheless impact the derived SFRs. As such, we assume a mean dust correction factor of $\times1.4$ at all redshift bins, as found by \citet{bouwens15}.

\begin{figure*}
\center
 \includegraphics[width=0.75\textwidth]{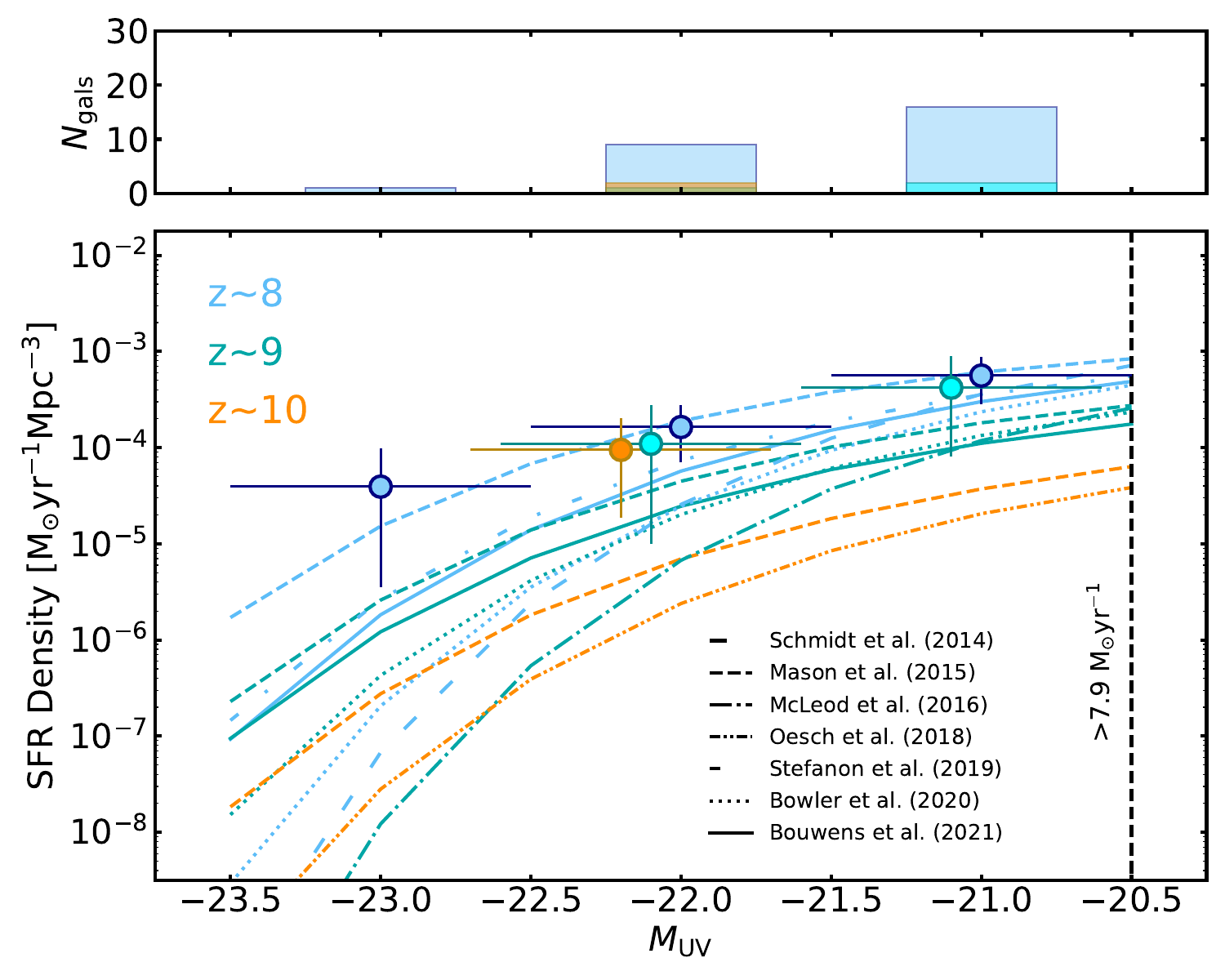}
 \caption{The bright end of the cosmic SFR density as a function of absolute \textit{UV} magnitude and redshift. Circles refer to the values derived in this analysis (horizontal error bars mark the magnitude bin edges) while dotted and dashed (colored) lines refer to integrations (over the same magnitude limits as the SuperBoRG data points) of LFs in the literature, assuming a consistent conversion factor. Blue colors refer to results at $z\sim8$, while cyan colors refer to $z\sim9$ results and orange colors to $z\sim10$ results (the $z\sim9$ and $z\sim10$ points are slightly offset in $M_{\rm UV}$ for clarity). The number of galaxies per redshift in each magnitude bin are presented as histograms in the top panel.}
 \label{fig:csfrd}
\end{figure*}

We plot the results of the above derivations in Figure~\ref{fig:csfrd}, where it becomes immediately clear that our results are in broad agreement with those inferred from LFs and a conversion factor. This is particularly true at the faintest end, where our $M_{\rm UV}\sim-21$ mag point is perfectly consistent with virtually all $z\sim8$ LF results. However, taking the results at face value, such consistency fades as one approaches brighter limits and an apparent tension arises: derivations with LFs can significantly underestimate the SFRD compared to values determined with the photometrically-derived SFRs. Such face-value discrepancies appear enhanced in higher-redshift derivations compared to lower-redshift derivations. However, the degree of such tension is challenging to quantify: small number statistics dominate the uncertainties, while LF-derived values can themselves differ by up to $>2$ dex depending on the assumed parameters. Additionally, an important caveat to consider are the different samples adopted in this study and by the various LF analyses, which are likely to introduce important differences themselves. We find that although our derived points are systematically higher than those from some LF-derived values (particularly at the brightest end), the differences are similar to those found between the LFs themselves (e.g., for $z\sim8$ derivations, we observe a $\sim0.5-1$ dex and $\sim0.5-2$ dex difference between our points and values derived using a $z\sim8$ LF from \citealt{bouwens21} and \citealt{schmidt14}, respectively, however the difference between the points of the two LFs themselves are $\sim0.5-1$ dex). Finally, as a sanity check, we also investigate the effects of magnification bias on our conclusions. While not expected to be significant among pure-parallel BoRG data sets \citep{mason15b}, several authors have found modest magnification effects for several luminous sources in large data sets ($\mu\approx1.2-1.5\times$; e.g., \citealt{calvi16,stefanon19}), warranting investigation. We thus assume a sample-wide magnification factor of $\mu=1.5$ and adjust the photometry of each of our $z\sim8-10$ sources accordingly, prior to re-fitting their photometry with our \texttt{Bagpipes} model and re-deriving SFRDs, in identical fashion to what is described in Section \ref{sec:sampleprops} and here above. We find that the magnification-corrected SFRDs resulting from the new $M_{\rm UV}$s and SFRs are virtually identical to their original values - with the exception of the removal of the brightest $z\sim9$ data point and addition of a $z\sim10$ point in the faintest $M_{\rm UV}$ bin, both of which display near identical offsets to the other $M_{\rm UV}$ data points. While this does not preclude some minor gravitational lensing effects on a source-by-source basis, we deem the observed offsets and our conclusions robust against magnification biases. Our findings thus point to two main conclusions, namely (i) that within uncertainties both SFRD values derived via SED-fitting and those derived with LFs and a conversion factor are broadly consistent with each other and (ii) the differences between the two approaches are dominated by the number counts of galaxies in each bin, pointing to cosmic variance effects and highlighting further the power and importance of (pure-)parallel observations, where such effects are minimized.

\section{Forecasts with JWST}
\label{sec:jwst}
Ultimately, \jwst\ will be required to confirm photometric galaxy candidates as well as the results of interpretations in the literature at $z>8$. While the facility's photometric capabilities alone will prove invaluable towards the search for high-$z$ galaxy candidates with (pure-)parallel observations (e.g., GO 2514, PI: Williams and GO 1571, PI: Malkan, as well as highlighted by the present and previous BoRG works) and yield order-of-magnitude improvements on current measurements (see e.g., the analysis by \citealt{rb21}), the advent of space-based NIR spectroscopy will ultimately prove to be the game-changer for galaxy evolution studies in the Epoch of Reionization. Low resolution ($R\sim100$) slit spectroscopy extending out to wavelengths of $\sim0-5 \mu$m and $\sim5-12 \mu$m with e.g., NIRSpec/Prism and MIRI/Slit will allow for simultaneous observations of the rest-frame \textit{UV} and optical, revealing emission line and continuum features in sub-hour exposures. For observations of Reionization-era galaxies at $z\sim8$, where NIR spectroscopy has been confined to largely unsuccessful observations of Ly$\alpha$ for secure redshifts, the availability of rest-frame optical lines such as [\OII]$\lambda\lambda$3726,3729 \AA, [\OIII]$\lambda\lambda$4959,5007 \AA\ and H$\beta$ will allow for independent redshift determinations as well as constraints on the state of the gas and ionizing mechanisms (e.g., sSFRs and metallicities). Furthermore, observations of the so-called ``Balmer break'' and stellar continuum at rest-frame optical wavelengths will provide accurate constraints on the ages, SFRs and stellar masses of such galaxies, something which has thus far been exclusive to objects with deep \spitzer/IRAC observations. Such observations and measurements form the basis for several accepted ERS, GTO and Cycle 1 observations and require exposures as low as $\sim30$ mins to obtain a S/N of $\gtrsim5\sigma$ over the majority of the rest-frame \textit{UV}-to-optical galaxy spectrum. As an illustration, we use the \texttt{Pandeia} Exposure Time Calculator tool to simulate a NIRSpec/Prism exposure of a luminous (SuperBoRG) galaxy inside the NIRSpec MSA. We select 0853$+$0310\_112 ($z_{\rm phot}=7.66$, $m_{\rm F160W}\sim25$ AB) as our show case example, and select a NRS readout pattern, 9 groups/integration, 1 integration/exposure, and 9 exposures (or dithers) total to simulate a $\sim1$ hr exposure. 

\begin{figure}[H]
\center
 \includegraphics[width=\columnwidth]{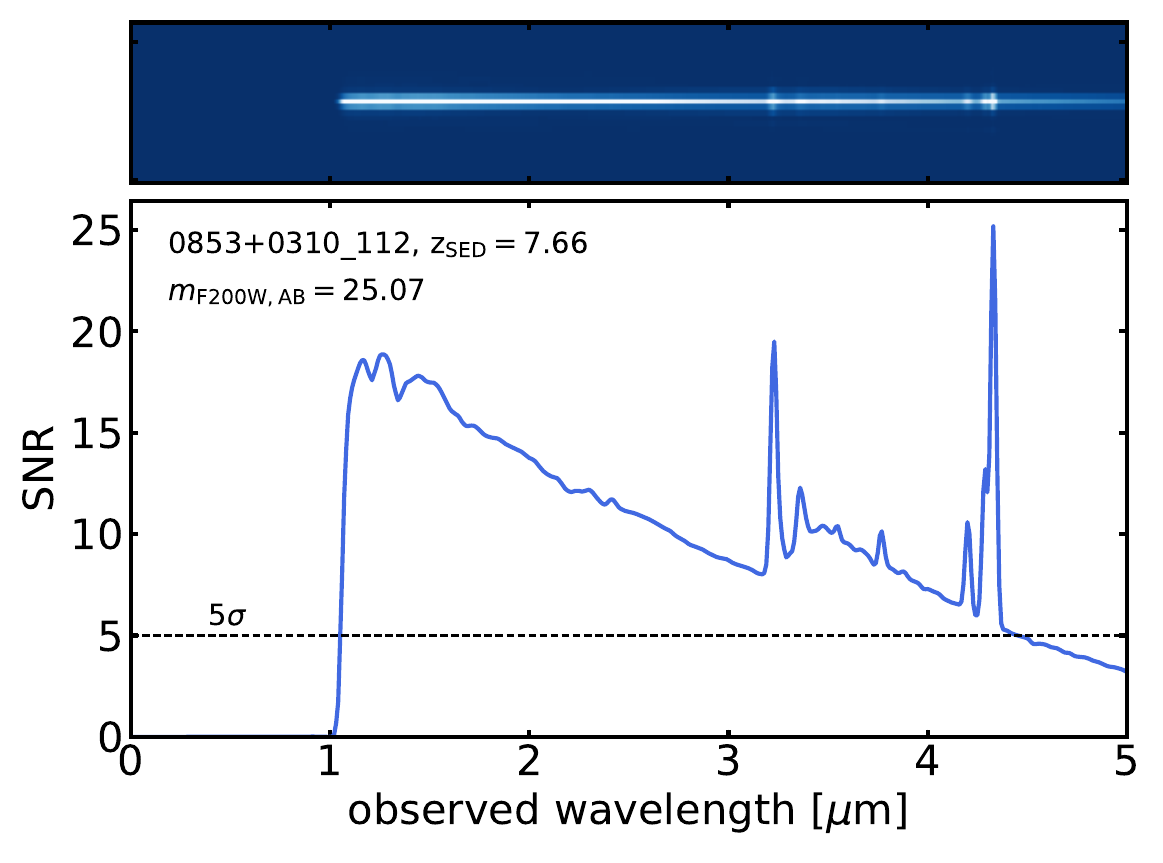}
 \caption{A simulated \jwst/NIRSpec (prism) spectrum of one of our selected $z\sim8$ sources using the best-fit \texttt{Bagpipes} SED and a 1 hr exposure resulting from 9 groups/integration, 1 integration/exposure, and 9 exposures total with an NRS readout pattern. The simulations indicate that the full rest-frame \textit{UV} and large portion of the rest-frame optical spectra are detected at $>5\sigma$ significance, including both continuum and emission line measurements.}
 \label{fig:nirspec}
\end{figure}

The resulting 2D and 1D spectra as a function of S/N are shown in Figure~\ref{fig:nirspec}, where the entire rest-frame \textit{UV} spectrum and blue half of the rest-frame optical spectrum are clearly detected ($>5\sigma$), including all continuum and emission line features. \jwst\ will undoubtedly prove a revolutionizing facility towards the analysis of high-$z$ galaxy evolution and $z\gtrsim8$ objects in particular: to this end and of special interest to the present work, 10 SuperBoRG galaxies are scheduled for observation and characterisation with NIRSpec/Prism spectroscopy as part of Cycle 1 observations (GO 1747, PI: Roberts-Borsani).

\section{Summary \& Conclusions}
\label{sec:conclusions}
We have conducted the largest space-based search to date for luminous $z\gtrsim8$ galaxy candidates using $\sim$1267 arcmin$^{2}$ of \hst\ (pure-)parallel observations from the SuperBoRG ACS+WFC3 compilation. We find 31 photo-$z$ galaxies over 29 unique sightlines with photometric redshifts of $z\sim8-10$ and investigate their rest-frame \textit{UV} and optical properties. Our main findings can be summarized as follows:

\begin{itemize}
\item Derived values of absolute magnitude and \textit{UV} continuum slope indicate our sample represent the brightest-end of the luminosity function with values spanning approximately $-21$ mag to $-23$ mag and relatively blue slopes of approximately $-1.5$ to $-2.5$, indicative of low-metallicity/dust systems in the early Universe. No significant trend in $\beta$ is found with either absolute magnitude or redshift, possibly illustrating the limiting dynamic range of the sample and/or the short evolutionary timescale between our selected redshifts.

\item A comparison of our sample's half-light radii (a median of 0.80$\pm$0.44 kpc) - derived from \hst/WFC3 detection images - with those presented in the literature clearly highlight the size-mass relation at high redshift and confirm that the most luminous and massive galaxies ($M_{\rm UV}<-21$ mag) display the largest sizes (i.e., a few kpc) while the faintest and least massive ($M_{\rm UV}>-17$ mag) galaxies display the smallest sizes (i.e., sub-kpc down to several pc).

\item For galaxies where \spitzer/IRAC observations are available, detailed SED-fitting reveal rest-frame optical properties that are similar to those of confirmed high-redshift galaxies: we observe stellar masses of log $M/M_{\odot}\sim9.5$, stellar ages of $\sim0-400$ Myrs, metallicities up to 0.7 $Z_{\odot}$ and low dust contents of $A_{\rm v}\sim0-1$ mag, pointing to a consistent physical picture for representative galaxies in the early Universe.

\item We select four BoRG sources for follow up $Y-$band spectroscopy with Keck/MOSFIRE to search for \lya\ emission. Two of those targets were selected as part of our primary sample, while the remaining two were selected as part of previous BoRG surveys. With $>3$ hrs of observations for each source, none display clear \lya\ emission, indicative of a robust selection function against low-redshift interlopers and of a predominantly neutral IGM at $z\sim8$.

\item The (pure)-parallel nature of our data sets allow us to perform a crucial and unbiased test of derivations of the the bright-end of the cosmic SFR density for comparison with derivations using luminosity functions and an uncertain conversion factor. Using detailed completeness simulations and SFRs from SED-fitting, we find consistency with LF results at the faintest end of our absolute magnitude range (i.e., $M_{\rm UV}=-21$ mag) but a possible underestimation of the CSFRD by LF-derived results at the brighter end ($M_{\rm UV}<-21$ mag). Such differences are, however, statistically consistent with the differences between derivations adopting various LF parameters, suggesting that discrepancies are unlikely to arise due to conversion factors but rather the number counts of galaxies upon which the LFs are constrained. Such an observation places even greater importance on the need for independent and unbiased pointings to search for luminous high-redshift galaxies.

\item 10 of our $z\sim8$ SuperBoRG sources have been selected for follow up prism spectroscopy with \jwst/NIRSpec as part of Cycle 1 observations (GO 1747, PI G. Roberts-Borsani). Simulations show 0.5-1 hr observations with the instrument will allow for $>5\sigma$ detections along the whole of the rest-frame \textit{UV} and majority of the rest-frame optical spectra. Such observations over (pure-)parallel-selected sources will go a long way towards gaining representative measurements of metallicities, specific-SFRs, stellar masses and stellar ages from continuum and emission line measurements previously unobtainable with current ground- and space-based facilities alone.
\end{itemize}

Constraints on galaxy evolution at redshifts beyond $z\gtrsim8$ have seen impressive progress thanks to large surveys undertaken with \hst\ and spectroscopic follow up of selected sources. However, the single-pointing nature of many such surveys raises the question of how representative resulting samples may be, particularly at the bright end of the luminosity function where clustering effects are expected to become important. (Pure-)parallel observations serve as an especially valuable alternative for searches of luminous $z\gtrsim8$ galaxies and, given the independent nature of the observations, provide a crucial benchmark for current and future interpretations of galaxy properties that are robust against cosmic variance. Ultimately, such results will require confirmation with upcoming facilities such as the \textit{James Webb} Space Telescope, however the large area afforded by the compilation of data sets in the SuperBoRG survey and results presented here offer a definitive legacy on the bright end of the luminosity function and a benchmark for future follow up with upcoming facilities.

\acknowledgments
We acknowledge support by NASA through grants HST-GO 16005, 15212, 14701, 13767, 12905, 12572, 11700, 15212.002, 15702.002 and HST-AR-15804.002-A from the Space Telescope Science Institute, which is operated by AURA, Inc., under NASA contract NAS 5-26555. We also acknowledge support from NSF through grant NSF-AST-1810822 ''COLLABORATIVE RESEARCH: The Final Frontier: Spectroscopic Probes of Galaxies at the Epoch of Reionization.''

The spectroscopic data presented herein were obtained at the W. M. Keck Observatory, which is operated as a scientific partnership among the California Institute of Technology, the University of California and the National Aeronautics and Space Administration. The Observatory was made possible by the generous financial support of the W. M. Keck Foundation. The authors also wish to recognize and acknowledge the very significant cultural role and reverence that the summit of Maunakea has always had within the indigenous Hawaiian community. We are most fortunate to have the opportunity to conduct observations from this mountain.

This research was supported in part by the Australian Research Council Centre of Excellence for All Sky Astrophysics in 3 Dimensions (ASTRO 3D), through project number CE170100013.

This research has benefitted from the SpeX Prism Library (and/or SpeX Prism Library Analysis Toolkit), maintained by Adam Burgasser at \url{http://www.browndwarfs.org/spexprism}.

We thank the anonymous referee for a thoughtful report which
strengthened this study.

\appendix

\section{Primary sample of $z\gtrsim8$ galaxies}
\label{sec:primesample}
Here below we tabulate (in Table~\ref{tab:candidates}) the main photometric colors and photo-$z$ properties of the final sample of high-confidence LBGs, after photometric-redshift selection, as well as their best-fit SEDs from \texttt{Bagpipes}.

\begin{table}[htp]
\centering
\resizebox{\textwidth}{!}{\begin{tabular}{lccccccccc}
\hline
\hline
Field\_ID & $\alpha_{\rm J2000}$ & $\delta_{\rm J2000}$ & $J_{\rm 125}$ & $H_{\rm 160}$ & $Y_{\rm 105}-J_{\rm 125}$ & $J_{\rm 125}-H_{\rm 160}$ & $z_{\rm phot}$ & $P(z>6.5)$ & $P_{\rm gal}$\\
 & [deg] & [deg] & [AB] & [AB] & [AB] & [AB] \\
\hline
\\
\multicolumn{10}{c}{$z\sim8$ high-confidence candidates} \\
\hline
0314$-$6712\_383 & 48.4497 & -67.2097 & 24.06$\pm$0.01 & 24.09$\pm$0.01 & $>$0.70 & -0.03$\pm$0.02 & 7.75$^{+0.20}_{-0.21}$ & 1.00 & 0.96 \\
0409$-$5317\_313 & 62.3371 & -53.2590 & 26.26$\pm$0.08 & 26.47$\pm$0.19 & $>$0.97 & -0.21$\pm$0.20 & 8.02$^{+0.23}_{-0.31}$ & 0.96 & 0.94 \\
0440$-$5244\_742 & 69.9457 & -52.7320 & 25.80$\pm$0.16 & 25.72$\pm$0.15 & --- & 0.09$\pm$0.22 & 7.84$^{+0.71}_{-0.11}$ & 1.00 & 0.99 \\
0830$+$6555\_244 & 127.5521 & 65.9035 & 25.60$\pm$0.09 & 25.32$\pm$0.14 & $>$0.98 & 0.28$\pm$0.17 & 7.75$^{+0.42}_{-0.29}$ & 0.90 & 0.92 \\
0853$+$0310\_112 & 133.1855 & 3.1467 & 24.97$\pm$0.04 & 25.07$\pm$0.08 & $>$0.65 & -0.09$\pm$0.09 & 7.66$^{+0.23}_{-0.23}$ & 1.00 & 0.93 \\
0925$+$1360\_899 & 141.2877 & 13.9991 & 25.69$\pm$0.13 & 25.71$\pm$0.25 & $>$0.69 & -0.02$\pm$0.28 & 7.58$^{+0.54}_{-0.20}$ & 0.74 & 0.97 \\
0948$+$5757\_697 & 147.0725 & 57.9549 & 25.84$\pm$0.10 & 25.63$\pm$0.16 & $>$0.81 & 0.20$\pm$0.19 & 7.66$^{+0.54}_{-0.25}$ & 0.72 & 0.82 \\
0955$+$4528\_914 & 148.8281 & 45.4895 & 26.12$\pm$0.11 & 25.85$\pm$0.20 & 1.79$\pm$0.11 & 0.28$\pm$0.23 & 8.20$^{+0.39}_{-0.57}$ & 0.90 & 0.93 \\
0956$+$2848\_986 & 149.0963 & 28.8095 & 26.80$\pm$0.12 & 26.44$\pm$0.19 & $>$1.63 & 0.36$\pm$0.22 & 8.11$^{+0.60}_{-0.19}$ & 1.00 & 1.00 \\
1017$-$2052\_310 & 154.3421 & -20.8697 & 26.00$\pm$0.16 & 25.73$\pm$0.21 & $>$1.52 & 0.27$\pm$0.27 & 8.11$^{+0.20}_{-0.65}$ & 0.75 & 0.94 \\
1033$+$5051\_164 & 158.1865 & 50.8416 & 25.91$\pm$0.07 & 26.03$\pm$0.24 & --- & -0.13$\pm$0.25 & 7.93$^{+0.52}_{-0.12}$ & 1.00 & 0.99 \\
1104$+$2813\_447 & 165.9720 & 28.2297 & 25.61$\pm$0.11 & 25.48$\pm$0.21 & $>$0.71 & 0.14$\pm$0.24 & 7.58$^{+0.61}_{-0.20}$ & 0.79 & 0.94 \\
1218$+$3008\_638 & 184.5579 & 30.1255 & 25.92$\pm$0.10 & 25.78$\pm$0.16 & $>$0.67 & 0.14$\pm$0.19 & 7.58$^{+0.46}_{-0.21}$ & 0.74 & 0.75 \\
1437$+$5043\_1241 & 219.2105 & 50.7260 & 25.80$\pm$0.03 & 25.65$\pm$0.08 & $>$0.88 & 0.15$\pm$0.08 & 7.84$^{+0.34}_{-0.15}$ & 1.00 & 1.00 \\
1515$-$1517\_698 & 228.7442 & -15.2704 & 25.87$\pm$0.11 & 25.97$\pm$0.24 & $>$0.91 & -0.10$\pm$0.26 & 7.66$^{+0.57}_{-0.22}$ & 0.84 & 0.58 \\
1558$+$0812\_601 & 239.5695 & 8.2083 & 25.10$\pm$0.10 & 25.00$\pm$0.11 & $>$0.90 & 0.10$\pm$0.15 & 7.66$^{+0.65}_{-0.17}$ & 0.99 & 0.55 \\
1917$-$3335\_929 & 289.3521 & -33.5924 & 26.11$\pm$0.15 & 25.93$\pm$0.26 & $>$1.12 & 0.18$\pm$0.30 & 8.11$^{+0.20}_{-0.64}$ & 0.80 & 0.81 \\
2203$+$1851\_1071 & 330.6929 & 18.8581 & 25.63$\pm$0.10 & 25.52$\pm$0.17 & --- & 0.11$\pm$0.20 & 7.84$^{+0.75}_{-0.14}$ & 1.00 & 0.99 \\
0728$+$0509\_232 & 111.9093 & 5.1312 & 25.25$\pm$0.05 & 24.98$\pm$0.10 & $>$1.47 & 0.27$\pm$0.11 & 8.20$^{+0.39}_{-0.45}$ & 1.00 & 0.96 \\
0104$+$0021\_339 & 15.8784 & 0.3370 & 25.34$\pm$0.08 & 25.59$\pm$0.21 & $>$0.67 & -0.25$\pm$0.23 & 7.66$^{+0.12}_{-0.30}$ & 0.70 & 0.70 \\
1149$+$2202\_169 & 177.1723 & 22.0172 & 25.54$\pm$0.13 & 25.84$\pm$0.20 & $>$0.60 & -0.30$\pm$0.24 & 7.66$^{+0.38}_{-0.25}$ & 0.87 & 0.65 \\
1149$+$2202\_343 & 177.1584 & 22.0221 & 25.59$\pm$0.18 & 25.00$\pm$0.12 & $>$1.21 & 0.60$\pm$0.22 & 8.38$^{+0.46}_{-0.24}$ & 0.92 & 0.82 \\
2134$-$0708\_2928 & 323.5509 & -7.1224 & 25.80$\pm$0.17 & 25.88$\pm$0.20 & $>$0.54 & -0.09$\pm$0.26 & 7.58$^{+0.47}_{-0.23}$ & 0.72 & 0.76 \\
0859$+$4114\_138 & 134.8251 & 41.2237 & 25.34$\pm$0.13 & 25.30$\pm$0.15 & $>$0.92 & 0.04$\pm$0.20 & 7.75$^{+0.46}_{-0.22}$ & 0.97 & 0.60 \\
0859$+$4114\_718 & 134.8402 & 41.2377 & 25.27$\pm$0.12 & 25.05$\pm$0.12 & $>$0.95 & 0.22$\pm$0.17 & 7.75$^{+0.42}_{-0.25}$ & 0.93 & 0.82 \\
1115$+$2548\_455 & 168.6620 & 25.8007 & $>$28.39 & 26.61$\pm$0.25 & $>$-0.12 & $>$1.78 & 8.38$^{+0.24}_{-0.13}$ & 0.99 & 0.99 \\
\hline
\\
\multicolumn{10}{c}{$z\sim9$ high-confidence candidates} \\
\hline
1607$+$1332\_996 & 241.7016 & 13.5468 & 26.15$\pm$0.21 & 25.38$\pm$0.13 & 1.72$\pm$0.21 & 0.77$\pm$0.25 & 8.76$^{+0.52}_{-0.34}$ & 0.97 & 1.00 \\
0037$-$3337\_563 & 9.3000 & -33.6243 & 26.39$\pm$0.21 & 25.96$\pm$0.17 & 1.86$\pm$0.21 & 0.43$\pm$0.27 & 8.76$^{+0.45}_{-0.40}$ & 1.00 & 0.96 \\
1420$+$3743\_1025 & 215.0595 & 37.7356 & 27.15$\pm$0.49 & 26.00$\pm$0.20 & $>$0.37 & 1.16$\pm$0.53 & 8.76$^{+0.49}_{-0.44}$ & 0.97 & 0.97 \\
\hline
\\
\multicolumn{10}{c}{$z\gtrsim10$ high-confidence candidates} \\
\hline
1459$+$7146\_344 & 224.7238 & 71.7814 & 27.84$\pm$0.60 & 25.91$\pm$0.12 & --- & 1.93$\pm$0.61 & 10.00$^{+0.46}_{-0.27}$ & 0.80 & 0.98 \\
1142$+$3020\_67 & 175.6406 & 30.3028 & 26.33$\pm$0.38 & 24.81$\pm$0.12 & 1.19$\pm$0.38 & 1.52$\pm$0.40 & 9.79$^{+0.52}_{-0.18}$ & 0.94 & 0.99 \\
\hline
\end{tabular}}
\caption{The final sample of high confidence $z\gtrsim8$ SuperBoRG galaxies selected based on color cuts and photometric redshifts as defined in Section \ref{sec:sampselect}.}
\label{tab:candidates}
\end{table}

\begin{figure*}
\center
 \includegraphics[width=0.9\textwidth]{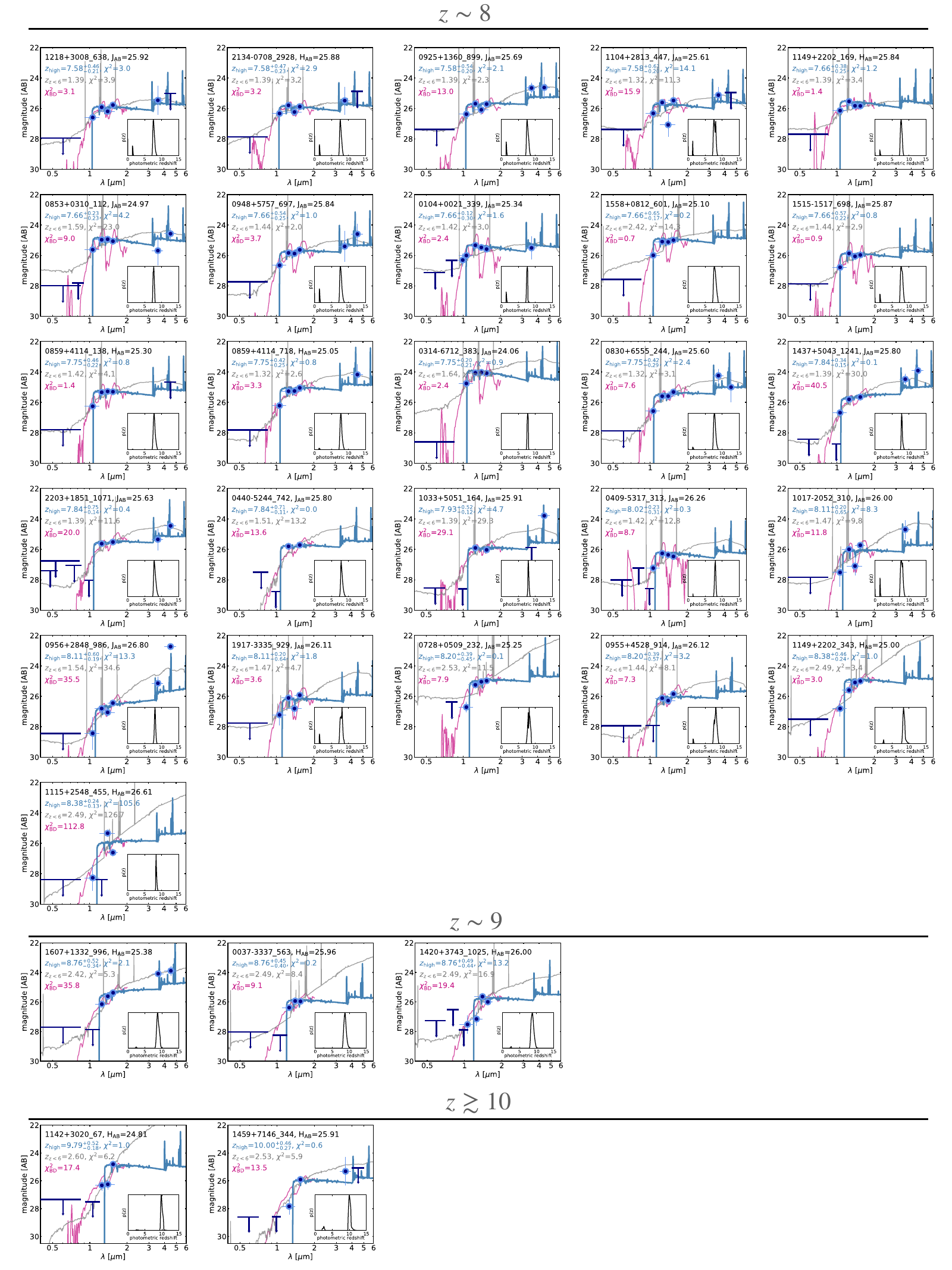}
 \caption{The best-fit SEDs and $P(z)$ from \texttt{Bapipes} and \eazy, respectively, for the galaxies tabulated in Table~\ref{tab:candidates}, along with their associated SuperBoRG photometry (blue points and upper limits). The best-fit high-$z$ SED is shown in blue while we also plot the (\eazy) best-fit SpeX brown dwarf template in magenta and a forced $z\sim1-4$ low-$z$ fit (gray) for illustration purposes only. All redshift and $\chi^{2}$ values are taken from \eazy\ fits but do not make use of the same prior. The figure in its entirety is available online.}
 \label{fig:seds}
\end{figure*}

\begin{figure*}
\center
 \includegraphics[width=\textwidth]{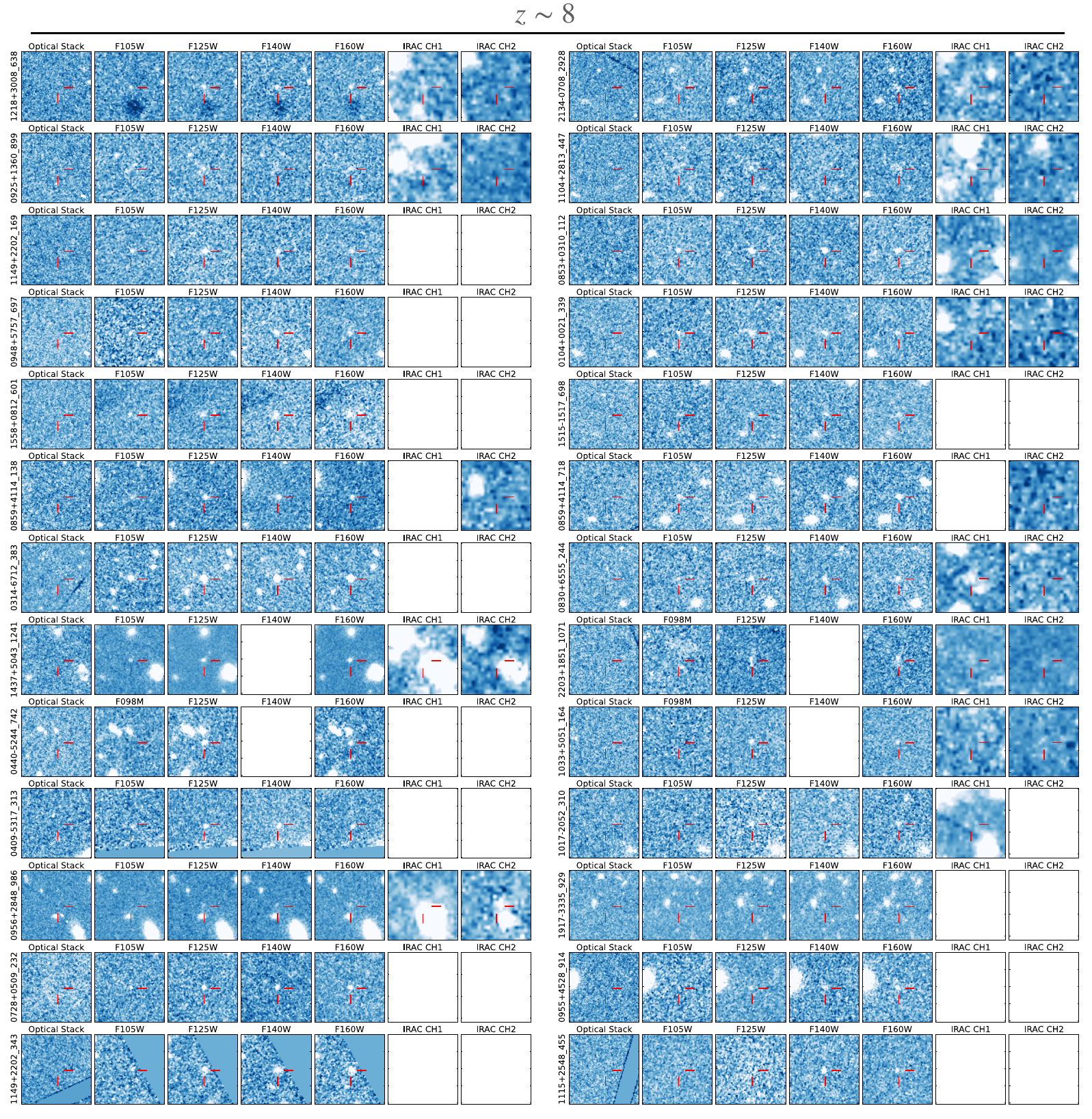}
 \caption{Postage stamp images of the SuperBoRG galaxies listed in Table \ref{tab:candidates}. From left to right, each set of images shows the optical stack of \hst\ filters blueward of the Lyman-break, F105W (or F098M, if F105W is unavailable), F125W, F140W, F160W, followed by the \spitzer/IRAC CH1 and CH2 bands. Each \hst\ image is a $6.5''\times6.5''$ cutout, while both \spitzer\ images are $12.8''\times12.8''$ cutouts.}
 \label{fig:postagestamps}
\end{figure*}


\begin{figure*}
\ContinuedFloat
\center
 \includegraphics[width=\textwidth]{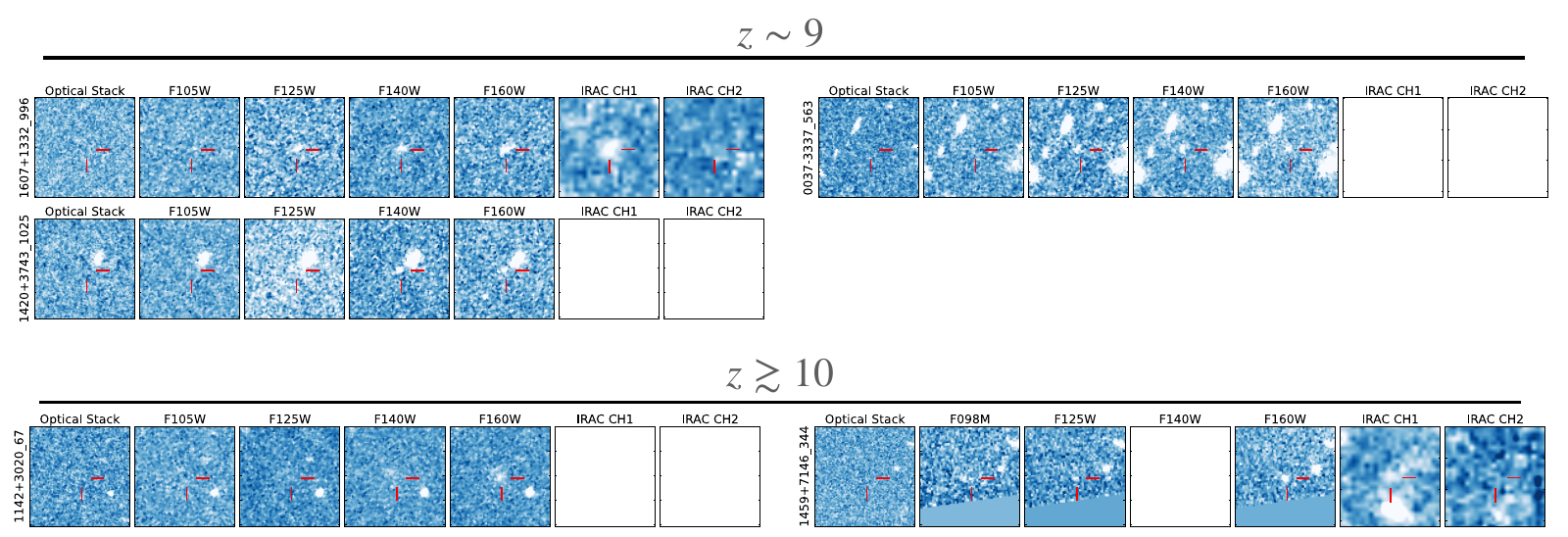}
 \caption[]{continued.}
\end{figure*}

\clearpage

\section{Flagged sample}
\label{sec:flagsample}
Similarly to Appendix \ref{sec:primesample}, here we tabulate (in Table~\ref{tab:candidates2}) the main NIR colors and photometric redshift properties of the final sample of flagged sources.

\begin{table}[htp]
\centering
\begin{tabular}{lccccccccc}
\hline
\hline
Field\_ID & $\alpha_{\rm J2000}$ & $\delta_{\rm J2000}$ & $J_{\rm 125}$ & $H_{\rm 160}$ & $Y_{\rm 105}-J_{\rm 125}$ & $J_{\rm 125}-H_{\rm 160}$ & $z_{\rm phot}$ & $P(z>6.5)$ & $P_{\rm gal}$\\
 & [deg] & [deg] & [AB] & [AB] & [AB] & [AB] \\
\hline
\\
\multicolumn{10}{c}{$z\sim8$ flagged candidates} \\
\hline
0037$-$3337\_669 & 9.2972 & -33.6185 & 25.79$\pm$0.09 & 25.75$\pm$0.17 & $>$0.58 & 0.04$\pm$0.20 & 7.58$^{+0.42}_{-0.20}$ & 0.76 & 0.40 \\
0143$+$4434\_56 & 25.8135 & 44.5440 & 25.45$\pm$0.09 & 25.11$\pm$0.12 & $>$1.05 & 0.34$\pm$0.15 & 8.29$^{+0.23}_{-0.60}$ & 0.92 & 0.67 \\
0751$+$2917\_337 & 117.7141 & 29.2715 & 26.57$\pm$0.06 & 26.45$\pm$0.15 & 1.77$\pm$0.06 & 0.13$\pm$0.16 & 8.11$^{+0.40}_{-0.26}$ & 1.00 & 0.98 \\
0830$+$6555\_840 & 127.5386 & 65.9185 & 25.28$\pm$0.09 & 25.30$\pm$0.18 & $>$1.20 & -0.02$\pm$0.20 & 8.11$^{+0.45}_{-0.21}$ & 1.00 & 1.00 \\
0859$+$4114\_81 & 134.8268 & 41.2229 & 25.08$\pm$0.07 & 24.81$\pm$0.12 & $>$1.03 & 0.26$\pm$0.14 & 7.84$^{+0.44}_{-0.33}$ & 1.00 & 1.00 \\
0925$+$1360\_131 & 141.3136 & 13.9867 & 25.76$\pm$0.10 & 25.95$\pm$0.24 & 2.14$\pm$0.10 & -0.19$\pm$0.26 & 8.11$^{+0.51}_{-0.19}$ & 0.99 & 0.98 \\
0926$+$4426\_1158 & 141.3865 & 44.4236 & 24.57$\pm$0.07 & 24.94$\pm$0.27 & --- & -0.37$\pm$0.28 & 7.75$^{+0.57}_{-0.19}$ & 1.00 & 0.91 \\
1048$+$4433\_687 & 162.1002 & 44.5675 & 25.81$\pm$0.14 & 25.39$\pm$0.19 & $>$1.63 & 0.41$\pm$0.23 & 7.93$^{+0.58}_{-0.41}$ & 0.95 & 0.70 \\
1152$+$3402\_693 & 177.9128 & 34.0445 & 25.97$\pm$0.13 & 25.92$\pm$0.23 & $>$1.04 & 0.05$\pm$0.26 & 7.66$^{+0.50}_{-0.24}$ & 0.79 & 0.83 \\
1456$+$2754\_381 & 224.0788 & 27.9013 & 25.78$\pm$0.12 & 25.88$\pm$0.27 & $>$1.00 & -0.10$\pm$0.29 & 7.75$^{+0.54}_{-0.28}$ & 0.84 & 0.55 \\
1558$+$0812\_368 & 239.5731 & 8.1985 & 25.65$\pm$0.11 & 25.63$\pm$0.12 & $>$0.54 & 0.03$\pm$0.17 & 7.66$^{+0.32}_{-0.27}$ & 0.84 & 0.52 \\
2008$-$6610\_1020 & 302.0242 & -66.1658 & 25.73$\pm$0.07 & 25.80$\pm$0.14 & $>$1.70 & -0.07$\pm$0.15 & 8.11$^{+0.33}_{-0.39}$ & 0.98 & 0.86 \\
2055$-$3408\_1026 & 313.6711 & -34.1323 & 25.47$\pm$0.03 & 25.56$\pm$0.11 & $>$0.59 & -0.10$\pm$0.11 & 7.66$^{+0.11}_{-0.28}$ & 0.93 & 0.57 \\
1048$+$1518\_72 & 161.9772 & 15.2852 & 25.27$\pm$0.15 & 25.33$\pm$0.19 & $>$1.38 & -0.06$\pm$0.25 & 8.11$^{+0.24}_{-0.52}$ & 0.97 & 0.91 \\
1313$+$1804\_2210 & 198.1954 & 18.0693 & 25.79$\pm$0.13 & 25.79$\pm$0.15 & $>$0.56 & -0.00$\pm$0.20 & 7.58$^{+0.43}_{-0.19}$ & 0.81 & 0.42 \\
1420$+$3743\_1078 & 215.0515 & 37.7338 & 24.80$\pm$0.15 & 24.41$\pm$0.17 & $>$1.82 & 0.39$\pm$0.23 & 8.11$^{+0.53}_{-0.37}$ & 1.00 & 0.95 \\
1456$+$2754\_665 & 224.0888 & 27.9029 & 25.76$\pm$0.13 & 25.49$\pm$0.12 & $>$1.54 & 0.27$\pm$0.17 & 8.11$^{+0.41}_{-0.49}$ & 0.99 & 0.90 \\
2212$-$0354\_102 & 333.0118 & -3.9183 & 25.72$\pm$0.17 & 25.93$\pm$0.22 & $>$0.61 & -0.22$\pm$0.27 & 7.58$^{+0.46}_{-0.19}$ & 0.84 & 0.90 \\
1416$+$1638\_150 & 213.9968 & 16.6134 & 25.67$\pm$0.05 & 26.45$\pm$0.19 & --- & -0.78$\pm$0.20 & 8.11$^{+0.14}_{-0.12}$ & 1.00 & 0.99 \\
\hline
\\
\multicolumn{10}{c}{$z\sim9$ flagged candidates} \\
\hline
1525$+$0960\_719 & 231.1971 & 9.9970 & 27.03$\pm$0.69 & 25.34$\pm$0.17 & 0.58$\pm$0.69 & 1.69$\pm$0.71 & 9.47$^{+0.35}_{-0.56}$ & 0.95 & 0.97 \\
\hline
\\
\multicolumn{10}{c}{$z\gtrsim10$ flagged candidates} \\
\hline
1442$-$0212\_245 & 220.5278 & -2.2053 & 27.84$\pm$0.95 & 25.85$\pm$0.16 & 0.33$\pm$0.95 & 2.00$\pm$0.97 & 10.45$^{+0.87}_{-0.50}$ & 0.80 & 0.98 \\
0409$-$5317\_291 & 62.2994 & -53.2907 & $>$27.96 & 25.59$\pm$0.16 & 0.07$\pm$0.00 & $>$2.37 & 9.89$^{+0.87}_{-0.21}$ & 0.92 & 0.99 \\
1014$-$0423\_166 & 153.5068 & -4.3776 & 25.87$\pm$0.30 & 24.05$\pm$0.07 & --- & 1.82$\pm$0.31 & 9.89$^{+0.47}_{-0.29}$ & 0.79 & 1.00 \\
1206$-$0852\_423 & 181.5962 & -8.8791 & 24.67$\pm$0.14 & 23.37$\pm$0.04 & --- & 1.30$\pm$0.15 & 9.89$^{+0.49}_{-0.24}$ & 0.92 & 1.00 \\
1413$+$0918\_1405 & 213.2092 & 9.3007 & $>$28.03 & 25.64$\pm$0.13 & 0.03$\pm$0.00 & $>$2.39 & 10.56$^{+0.47}_{-0.46}$ & 1.00 & 0.99 \\
1031$+$5052\_602 & 157.6526 & 50.8737 & $>$28.74 & 24.72$\pm$0.12 & --- & $>$4.02 & 10.56$^{+0.57}_{-0.06}$ & 1.00 & 1.00 \\
1437$+$5043\_259 & 219.2511 & 50.7144 & $>$28.99 & 25.95$\pm$0.18 & --- & $>$3.04 & 10.56$^{+0.60}_{-0.25}$ & 1.00 & 0.99 \\
1237$+$2544\_806 & 189.3749 & 25.7427 & $>$28.41 & 25.81$\pm$0.12 & 0.05$\pm$0.00 & $>$2.60 & 12.16$^{+0.28}_{-0.25}$ & 1.00 & 1.00 \\
1142$+$2647\_1280 & 175.4986 & 26.7807 & $>$28.19 & 25.91$\pm$0.14 & 0.09$\pm$0.00 & $>$2.27 & 12.16$^{+0.18}_{-1.24}$ & 1.00 & 0.99 \\
\hline
\end{tabular}
\caption{The final sample of flagged $z\gtrsim8$ SuperBoRG galaxies, selected based on color cuts and photometric redshifts as defined in Section \ref{sec:sampselect} but flagged during visual inspection. The table adopts the same format as Table~\ref{tab:candidates}.}
\label{tab:candidates2}
\end{table}

\clearpage

\section{Comparison of Dropouts Galaxies with Previous BoRG Surveys}
\label{sec:compare}
Given the SuperBoRG survey is largely comprised of previous (and new) BoRG data sets, it is informative to compare our dropout sources to the primary samples of galaxy candidates found in the most recent $z\sim8$ and $z\sim9-10$ BoRG analyses, presented in \citet{trenti11}, \citet{bradley12}, \citet{schmidt14}, \citet{calvi16}, \citet{livermore18} and \citet{morishita18} (hereafter T11, B12, S14, C16, L18 and M18 respectively). The comparison is separated out into sources detected in WFC3/F125W imaging (T11, B12 and S14) and those detected in WFC3/F140W+F160W (C16, L18, M18), for convenience.

\subsection{F125W-detected, $z\sim8$ LBGs}
We begin by examining the $z\sim8$ $Y_{098}$ dropout sources found by B12 and S14, who analyze 59 and 14 unique fields from the BoRG09+BoRG12 and BoRG13 data sets, respectively, spanning an effective area of $\sim$213 arcmin$^{2}$ and $\sim$40 arcmin$^{2}$. Two fields from the S14 study overlap with those analyzed in B12, with the former study presenting updated photometry and/or additional imaging to expand the $Y-$dropout search area. Thus, in the case of overlap, we adopt the sources and nomenclature of S14. Additionally, in the case of B12, the study also includes a re-analysis and expansion of the T11 data set and thus serves as an update from that study. From the searched BoRG fields, B12 and S14 find a total of 33 and 11 LBGs in 23 and 7 unique fields. Accounting for overlap (two sources), the total sample is comprised of 42 candidates in 28 unique fields ($\sim$123 arcmin$^{2}$, assuming 4.4 arcmin$^{2}$ per field).

We begin by exploring whether the same sources are detected within 1$''$ of their central coordinates in the improved SuperBoRG $J_{\rm 125}$ images. All but 5 sources (borg\_1153$+$0056\_540, borg\_1437$+$5043\_879, borg\_1632$+$3733\_694, borg\_2132$+$1004\_24 from B12 and BoRG\_1358$+$4334\_482 from S14) are detected in our images. Those galaxies which are not detected had reported $J-$band magnitudes of 26.5-27.4 AB, placing them amongst some of the faintest of the respective samples. Considering the improved noise constraints in the SuperBoRG images, the aforementioned galaxies are likely to have been false-positive detections.

Next we check how many of those detected sources are picked up as dropout galaxies using our updated photometry and dropout criteria. The NIR color cuts adopted by both the B12 and S14 analyses are nearly identical to those used here (and identical between themselves), while there are some differences in the S/N requirements and the optical non-detection filters used. For the NIR color cuts, this work adopts identical cuts to B12 and S14 but adopts an additional requirement of $J_{\rm 125}-H_{\rm 160}<0.5$. For the S/N requirements and non-detection filters, both B12 and S14 require S/N$_{J}\geqslant5.0$ and S/N$_{H}\geqslant2.5$ (these increase to S/N$_{J}\geqslant8.0$ and S/N$_{H}\geqslant3.0$ in the case of B12 for their high confidence sample) and S/N$_{V}\leqslant1.5$ in the $V-$band (typically F606W or F600LP) optical filter, while this work adopts more stringent S/N cuts of S/N$_{J}\geqslant6$, S/N$_{H}\geqslant4.0$, and S/N$_{\rm optical}<1.0$ for \textit{all} available filters blueward of (and including) $I_{\rm 814}$. Considering the above, of the 37 LBGs detected here, this work re-selects only 7 LBGs - 6 of these (borg\_0440$-$5244\_682, borg\_0751$+$2917\_229, borg\_1033$+$5051\_126, borg\_2203$+$1851\_1061, borg\_1131$+$3114\_1244, borg\_1437$+$5043\_172) derive from B12 and 1 from S14 (BoRG\_1437$+$5043\_r2\_637, also identified in B12).

\begin{figure}
\center
 \includegraphics[width=\columnwidth]{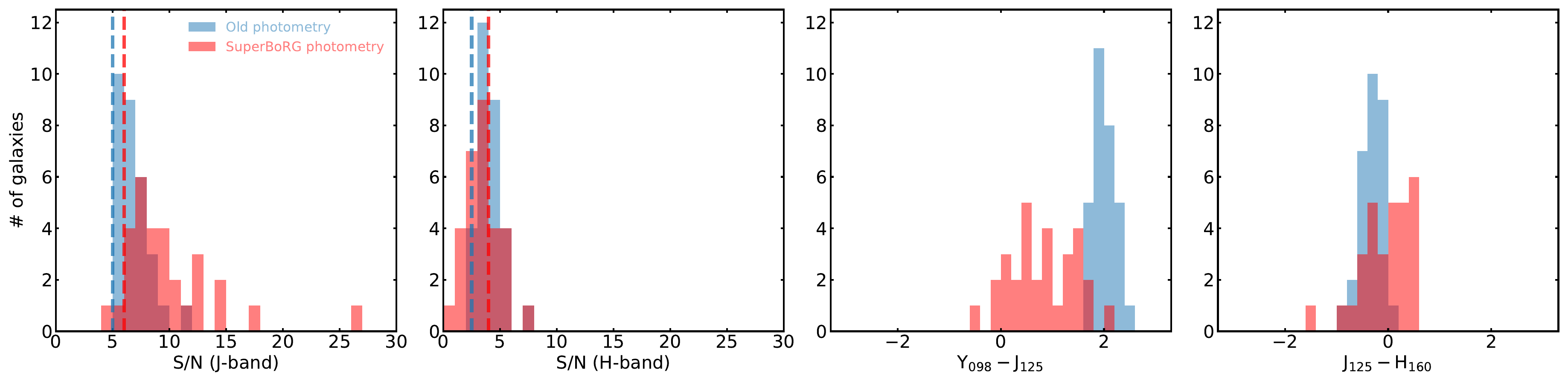}
 \caption{The (isophotal) photometry from the combined B12 and S14 galaxy samples (blue) compared to the (aperture) photometry adopted by SuperBoRG, for the same galaxy candidates. A significant shift towards bluer $Y_{\rm 098}-J_{\rm 125}$ colors is found in the latter photometry, suggesting previous photometry may have been overestimating the number of high-$z$ galaxy candidates based on artificially red dropout colors. The dashed lines highlight the S/N threshold required in the selection of candidates using the old photometry \citep{bradley12,schmidt14} and the new photometry presented here.}
 \label{fig:photocomp}
\end{figure}

\begin{figure}
\center
 \includegraphics[width=0.5\columnwidth]{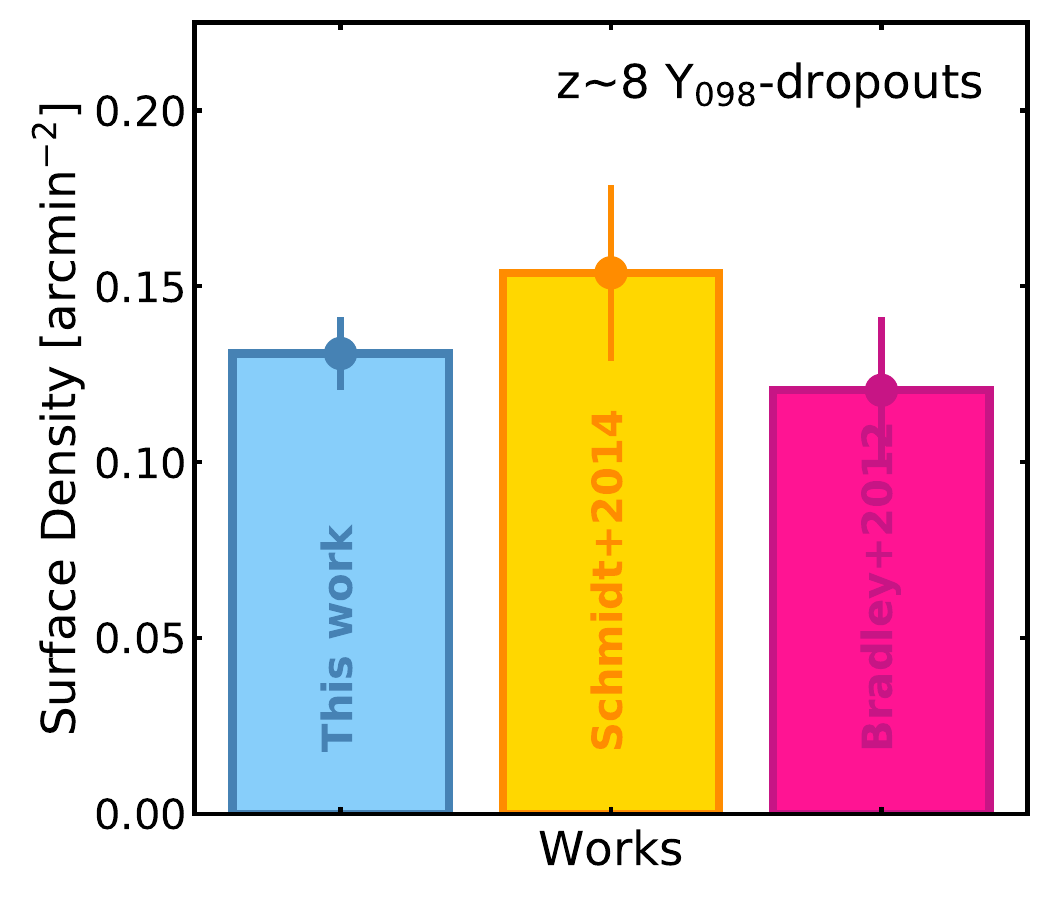}
 \caption{The number density of $z\sim8$ dropout candidates found by this work (blue, including both high-confidence and flagged candidates), \citet{schmidt14} (orange) and \citet{bradley12} (pink). All three studies make use of the $Y_{\rm 098}$ filter for NIR color selection. The numbers assumed here are 163 candidates over 1245.2 arcmin$^{2}$ for SuperBoRG, 38 candidates over 247 arcmin$^{2}$ for \citet{schmidt14} and 33 candidates over 274 arcmin$^{2}$ for \citet{bradley12}.}
 \label{fig:densityplot}
\end{figure}

Clearly the large majority of the detected galaxies in the combined B12 and S14 LBG sample are not selected as dropouts here (30/37), and we attribute this to two primary factors. Adopting the new photometry from SuperBoRG, the first factor is our more stringent $H-$band S/N threshold, which 21 of the 30 rejected B12+S14 dropouts fail. The second is the systematic shift towards bluer $Y_{\rm 098}-J_{\rm 125}$ colors, which results in 28 galaxies failing the $Y_{\rm 098}-J_{\rm 125}>$1.75 threshold. Such an effect is not completely unexpected, since in performing a comparison between isophotal (adopted by B12 and S14) and aperture (adopted by SuperBoRG) photometry \citet{morishita21} found that for sources fainter than $\sim$24 AB isophotal photometry produced systematically redder colors compared to aperture photometric measurements, implying that the use of the former can lead to significantly larger samples of false-positive high-$z$ dropouts. This is highlighted in Figure~\ref{fig:photocomp}, where we compare the photometry adopted by B12+S14 and the refined photometry from SuperBoRG, for the B12+S14 sample of galaxies discussed above. Thus, considering the faint nature of many of the previously-identified sources and the improved aperture photometry that SuperBoRG affords, we conclude that the majority of the sources were rejected as false-positive candidates. Finally, returning to the 7 dropout candidates re-identified in  this work, 2 of those are rejected based on visual inspection (borg\_1131$+$3114\_1244 and borg\_1437$+$5043\_172 from B12) and the remaining 5 candidates are included in the final photo-$z$ catalog (we note that borg\_0751$+$2917\_229 is flagged in our visual inspection). While clearly we observe a much smaller number of $z\sim8$ galaxies in the fields studied by S14/B12, we note that our search for $Y_{\rm 098}$ dropouts over our entire SuperBoRG data set yields a similar number density to each of the two aforementioned studies ($\sim0.125-0.15$ arcmin$^{-2}$; see Figure~\ref{fig:densityplot}), with differences consistent with cosmic variance and noise.

\subsection{F140W+F160W-detected, $z\sim9-10$ LBGs}
Using data from the BoRG[$z9-10$] survey (\hst/WFC3 observations over 28 unique fields, spanning a effective search area of $\sim$130 arcmin$^{2}$), C16 performed a search for $z\sim8-10$ LBGs using a F140W+F160W detection image and identical NIR color cuts to those utilized in this work (adopting the $Y_{\rm 105}$ filter for $z\sim8$ candidates), resulting in a sample of 9 LBGs at $z\gtrsim8-10$. L18 followed up 2 of these candidates with $Y_{\rm 098}$ \hst/WFC3 imaging and concluded one of these was likely to be a $z\sim2$ interloper, thus discarding it from the $z>8$ sample and bringing the total down to 8 high-$z$ candidates. Finally, M18 extended and updated the $z\sim9-10$ portion of the analysis (M18 did not search for $z\sim8$ candidates) to 79 independent fields, rejecting both $z\sim10$ candidates from C16 and one of the $z\sim9$ candidates (the same as was rejected by L18), however in the process finding an additional and previously unidentified $z\sim10$ candidate. Thus, the final C16 sample was reduced to two $z\sim9$ galaxies and four $z\sim8$ galaxies, with an additional $z\sim9$ candidate from M18. Similarly to the previous section, we adopt the results and nomenclature of the latest study where applicable (namely, M18).

Once again we begin by exploring whether the most up-to-date candidates (i.e., the four $z\sim8$ candidates from C16 - 0116$+$1425\_747, 0853$+$0310\_145, 1103$+$2913\_1216 and 1152$+$3402\_912 - and three updated $z\sim9$ candidates in M18 - 2140$+$0241$-$303, 0956$+$2848$-$98 and 2229$-$0945$-$394) are detected in our $JH+H$ detection image and find all of them are once again detected. Continuing to the color selection of the candidates, we find a total of three of the candidates are re-selected using SuperBoRG photometry: two of the $z\sim8$ candidates from C16 (0853$+$0310\_145 and 1103$+$2913\_1216) and one $z\sim9$ candidate from M18 (2229$-$0945$-$394). Additionally, 2229$-$0945$-$394 is rejected upon visual inspection due to non-negligible flux contribution in optical bands. Of the four color-rejected candidates, one fails our $JH-$band S/N threshold, while the remaining three all fail our $Y_{\rm 105}-JH<0.3$ criterium in similar fashion to the B12+S14 $z\sim8$ sources due to the bluer colors observed from aperture photometry. Finally, of the two remaining candidates that we re-identify in this work, 0853$+$0310\_145 makes it into the final photo-$z$ catalog as a $z\sim8$ galaxy, while 1103$+$2913\_1216 is identified as a $z_{\rm phot}=7.75$ galaxy however is not included in the final catalog as it has $P(z>6.5)=0.12$. We also note that we previously identified 0853$+$0310\_145 in \citet{morishita20} as a potential QSO using SuperBoRG photometry: the source was observed with Keck/MOSFIRE for a total of $\sim$6.6 hrs in $Y-$band, but did not reveal Ly$\alpha$ emission.

\clearpage

\section{Skew-normal priors used for photometric redshift estimation with \texttt{EAzY}}
We tabulate here (Table~\ref{tab:prior}) the double Skew-normal profiles that were constructed and described in Section \ref{subsubsec:prior}. The profiles are separated as a function of apparent $H_{\rm 160}$ magnitude and dropout redshift and fed into \texttt{EAzY} for each catalog according to their dropout criteria.

\begin{table*}[htp]
    \centering
    \begin{tabular}{lcccccccccc}
        \hline
        \hline
        $H_{\rm 160}$ & $A_{1}$ & $\mu_{1}$ & $\sigma_{1}$ & $\alpha_{1}$ & $A_{2}$ & $\mu_{2}$ & $\sigma_{2}$ & $\alpha_{2}$ & $f_{z<7}$ & $N_{\rm gals}$ \\
        \hline
        \\
        \hline
        \multicolumn{11}{c}{$Y_{\rm 098+105}\,J_{\rm 125}$-dropouts} \\
        \hline
        23.00 & 121.07 & 2.47 & 0.13 & 0.00 & 269.60 & 7.45 & 0.91 & 4.85 & 0.91 & 3153 \\
        23.50 & 399.00 & 2.50 & 0.10 & 0.07 & 918.55 & 7.41 & 0.88 & 6.37 & 0.06 & 10287 \\
        24.00 & 1042.50 & 2.47 & 0.11 & 0.39 & 2755.33 & 7.39 & 0.86 & 9.30 & 0.05 & 30320 \\
        24.50 & 2128.50 & 2.48 & 0.10 & 0.38 & 7133.77 & 7.38 & 0.81 & 8.25 & 0.05 & 73729 \\
        25.00 & 2131.74 & 2.52 & 0.10 & 0.00 & 15350.45 & 7.36 & 0.77 & 7.94 & 0.04 & 149756 \\
        25.50 & 2871.22 & 1.26 & 0.23 & 8.07 & 28769.51 & 7.35 & 0.76 & 7.92 & 0.02 & 281214 \\
        26.00 & 28376.56 & 1.39 & 0.14 & 9.96 & 45892.18 & 7.37 & 0.59 & 5.83 & 0.04 & 397724 \\
        26.50 & 23527.49 & 1.44 & 0.11 & 1.03 & 1335.96 & 7.84 & 0.22 & 0.00 & 0.16 & 37001 \\
        \hline
        \multicolumn{11}{c}{$Y_{\rm 105}\,JH_{\rm 140}$-dropouts} \\
        \hline
        23.00 & 134.00 & 2.47 & 0.13 & 0.00 & 106.15 & 8.18 & 0.86 & 6.10 & 0.91 & 1308 \\
        23.50 & 456.44 & 2.49 & 0.11 & 0.00 & 308.83 & 8.17 & 0.85 & 6.36 & 0.16 & 3823 \\
        24.00 & 1268.64 & 2.49 & 0.10 & 0.00 & 854.85 & 8.15 & 0.79 & 5.66 & 0.17 & 9862 \\
        24.50 & 2232.00 & 2.45 & 0.11 & 0.80 & 1895.28 & 8.13 & 0.76 & 5.85 & 0.16 & 20855 \\
        25.00 & 2356.50 & 2.48 & 0.10 & 0.38 & 3562.22 & 8.12 & 0.75 & 5.78 & 0.15 & 36061 \\
        25.50 & 1602.81 & 2.49 & 0.13 & 0.00 & 6246.85 & 8.13 & 0.74 & 5.97 & 0.09 & 60147 \\
        26.00 & 3734.32 & 2.46 & 0.11 & 0.00 & 9546.13 & 8.15 & 0.60 & 5.10 & 0.05 & 77231 \\
        26.50 & 1996.50 & 2.44 & 0.10 & 1.07 & 15.48 & 8.25 & 0.15 & 4.20 & 0.07 & 2534 \\
        \hline
        \multicolumn{11}{c}{$J_{\rm 125}\,H_{\rm 160}$-dropouts} \\
        \hline
        23.00 & 0.48 & 2.01 & 1.14 & 340.17 & 60.41 & 9.68 & 0.82 & 5.21 & 0.66 & 622 \\
        23.50 & 6.83 & 3.01 & 0.35 & 0.00 & 116.85 & 9.61 & 0.86 & 8.04 & 0.01 & 1306 \\
        24.00 & 21.56 & 2.90 & 0.51 & 0.00 & 237.53 & 9.59 & 0.83 & 6.50 & 0.02 & 2647 \\
        24.50 & 57.71 & 2.00 & 0.93 & 3042198.68 & 425.18 & 9.58 & 0.83 & 7.76 & 0.05 & 5137 \\
        25.00 & 218.48 & 2.00 & 0.91 & 869031.74 & 727.78 & 9.62 & 0.80 & 7.82 & 0.12 & 9754 \\
        25.50 & 821.16 & 2.00 & 1.17 & 264146.03 & 1230.74 & 9.61 & 0.79 & 6.85 & 0.23 & 23382 \\
        26.00 & 1314.85 & 2.00 & 1.10 & 103740.20 & 851.75 & 9.60 & 0.80 & 25307.75 & 0.46 & 25380 \\
        \hline 
    \end{tabular}
    \caption{The empirical double Skew-normal prior derived using simulated \hst\ ACS and WFC3 photometry from EGG over 10000 realizations of a 0.1 deg$^{2}$ field, and applying our dropout selection criteria described in Sections \ref{sec:colors} and \ref{subsubsec:prior}. Magnitude bins are 0.5 mag wide, while the redshift grid used for fitting ranged from $0<z<15$ in steps of $\delta z=0.2$. The final two columns indicate the fraction of low-$z$ ($z<7$) interlopers and the total number of dropouts in each magnitude-redshift bin, respectively. For galaxies outside of the limits of the magnitude ranges, we adopt the Gaussian parameters of the first/last row of each sample.}
    \label{tab:prior}
\end{table*}

\end{document}